\documentclass[
    10pt,
    twocolumn,
    secnumarabic,
    superscriptaddress,
    nobibnotes,
    nofootinbib,
    showkeys,
    aps,
    prd
]{revtex4-2}

\usepackage[T1]{fontenc}       
\usepackage{amsmath, amssymb}  
\usepackage{amsfonts}          
\usepackage{graphicx}          
\usepackage{bm}                
\usepackage{xcolor}            
\usepackage{comment}           
\usepackage{hyperref}          
\usepackage{hyperref}

\providecommand{\U}[1]{\protect\rule{.1in}{.1in}}

\newcommand{\be}{\begin{equation}}
\newcommand{\ee}{\end{equation}}

\usepackage{multirow}

\newcommand{\mincir}{\raise
-3.truept\hbox{\rlap{\hbox{$\sim$}}\raise4.truept\hbox{$<$}\ }}
\newcommand{\magcir}{\raise
-3.truept\hbox{\rlap{\hbox{$\sim$}}\raise4.truept\hbox{$>$}\ }}

\usepackage{enumitem}

\begin{document}

\title{ Conformally Coupled Scalar-Field Dark Energy: Cosmological Dynamics with Early- and Late-Measurements}
\author{Shambel Sahlu}
\email{shambel.sahlu@nithecs.ac.za} 
\affiliation{Centre for Space Research, North-West University, Potchefstroom 2520, South Africa}
 \affiliation{National Institute for Theoretical and Computational Sciences (NITheCS), South Africa}
\author{Abunie Gezahegn}
\email{gabunie21@gmail.com } 
\affiliation{Entoto Observatory and Research Center, Space Science and Geospatial Institute, Ethiopia}
\author{Amare Abebe}
\email{amare.abebe@nithecs.ac.za} 
\affiliation{Centre for Space Research, North-West University, Potchefstroom 2520, South Africa}
 \affiliation{National Institute for Theoretical and Computational Sciences (NITheCS), South Africa}
 \author{Gonzalo J. Olmo} \email{gonzalo.olmo@uv.es }
\affiliation{Instituto de Física Corpuscular (IFIC), CSIC‐Universitat de València, Spain}
 \author{Diego Rubiera-Garcia }
\email{drubiera@ucm.es}
\affiliation{Departamento de F\'isica Te\'orica and IPARCOS, Universidad  Complutense de Madrid,   Spain}

\vspace{2cm}
\begin{abstract}
In this work, the cosmology of the conformal interacting dark energy model is investigated by introducing a scalar field representing dark energy (DE) coupled to dark matter (DM) via a conformal transformation.  This coupling yields a class of scalar-tensor theories; the interaction term ${Q}$, which indicates the flow of energy between the dark sectors, is proportional to the trace of the energy–momentum tensor of the DM fluid, $T^{\mathrm{DM}}_{\mu\nu}$,  as ${Q}=\frac{C'(\phi)}{2C(\phi)}g^{\mu\nu}T^{\mathrm{DM}}_{\mu\nu}$, where $C(\phi)$ is the conformal function of a scalar potential (coupling) field. We assume the power-law parametrization $C(\phi)\propto (1+\phi)^m$, with $m$ as the coupling parameter, and determine the direction and magnitude of energy flow between the DM and DE scalar-field components. 
We model DE in three scenarios: conformally interacting dark energy (CIDE) coupled to a scalar field, uncoupled scalar-field DE (UCSFDE), and standard $ \Lambda$CDM models. We confront these models with cosmological measurements to test the viability of CIDE and UCSFED against $\Lambda$CDM and constrain them using a combination of early- and late-time data, namely: Cosmic Microwave Background measurements from Planck 2018 Legacy (\texttt{CMB});  Baryon Acoustic Oscillation (BAO) data from the Dark Energy Survey (\texttt{  DESI DR2 DR2 BAO}); and Supernova Type Ia distance compilations (\texttt{PantheonPlus, PantheonPlus + SH0ES, Union3}, and \texttt{DES-Dovekie}).  Parameter inference is performed with  Monte Carlo Markov Chain (\texttt{MCMC}) simulations using \texttt{COBAYA} and a modified \texttt{CLASS}. Further statistical analysis by Akaike information criterion (AIC) and the Bayesian information criterion (BIC) is summarized to assess the viability of the model in comparison with the standard cosmological model, investigating the model's potential to alleviate the $H_0$ and $S_8$ tensions.  We present numerical results for background and perturbative thermodynamic quantities, namely the matter power spectrum, the CMB anisotropy power spectrum, the transfer function, the growth factor, and redshift-space distortions, to evaluate the model for structure growth studies. Our results support conformally coupled dark-source models as a robust statistical alternative to $\Lambda$CDM.
\end{abstract} 
\date{\today}
\maketitle
\section{Introduction}
\noindent

The $\Lambda$CDM model has been successful in accommodating many cosmological data sets, including the universe's accelerating expansion \cite{riess1998observational,perivolaropoulos2008six}, cosmic microwave background (CMB) anisotropies \cite{aghanim2020planck},  large-scale structures of the universe \cite{bernardeau2002large, bull2016beyond}, and the observed abundances of various types of light nuclei such as hydrogen, deuterium, helium, and lithium  \cite{zwicky1933red, schramm1998big}. Despite its success, the $\Lambda$CDM model still faces significant challenges \cite{amendola2010dark,clifton2012modified,bull2015beyond,di2021snowmass2021}, notably the nature and magnitude of the cosmological constant, the cosmic coincidence problem, the $H_0$ tension, and the full dynamics of dark sources; see Refs. \cite{di2021realm,di2026tensions,dainotti2025new,navone2026creation,dainotti2024new,dainotti2022evolution}.  In the $\Lambda$CDM model,  a cosmological constant $\Lambda$ is responsible for driving the cosmic acceleration, while dark matter (DM) and dark energy (DE) evolve independently. Beyond $\Lambda$CDM, alternative DE models have been proposed in the literature in which interactions between the dark sectors are present, involving the transfer of energy between DM and DE \cite{amendola2000coupled,clifton2012modified, sahlu2026observational,vanderWesthuizen:2025vcb,van2025compartmentalization,sahlu2026diffusive,sahlu2026testing,
vanderWesthuizen:2025mnw,vanderWesthuizen:2025rip}. This mechanism provides effective solutions to several cosmological problems, and the flow of energy between DM and DE directly affects cosmic evolution \cite{Baldi:2008ay,Tocchini-Valentini:2001wmi,Chimento:2009, SantanaJunior:2024cug, Chimento:2012aea, Pan:2015,
Sharov:2017iue, Yang:2019, Paliathanasis:2019hbi, Leon:2020pvt, Okengo:2024mub, Valiviita:2008iv, Caldera-Cabral:2008yyo,
vanderWesthuizen:2025vcb,
vanderWesthuizen:2025rip,silva2026one,pu2015early,murgia2016constraints,zhai2023consistent,lucca2021multi,wang2016dark,DiValentino:2017iww}. The transfer of energy between DM and DE in an interacting scenario can be described via at least two distinct  approaches: 

i)  Phenomenological interacting DE (IDE) model \cite{Baldi:2008ay,Tocchini-Valentini:2001wmi, SantanaJunior:2024cug,Chimento:2012aea,Sharov:2017iue, Paliathanasis:2019hbi,paliathanasis2025compartmentalization,
Leon:2020pvt,Okengo:2024mub,Valiviita:2008iv, Caldera-Cabral:2008yyo,
vanderWesthuizen:2025vcb,
vanderWesthuizen:2025mnw,vanderWesthuizen:2025rip, pu2015early,murgia2016constraints,zhai2023consistent,lucca2021multi,wang2016dark,DiValentino:2017iww}, which incorporates the exchange of energy-momentum by introducing the ad hoc interacting term $Q^\mu$ directly into the fluid conservation equations, typically parameterized as a function of the background expansion rate $H(a)$ and dark sector densities $\rho_{i}$, which means  $Q\propto H \rho_{DM}$,  or  $ Q\propto  H\rho_{de}$, etc. Indeed, this approach is mathematically convenient for constraining the energy transfer. However,  it lacks a fundamental Lagrangian foundation, rendering cosmological perturbation theories vulnerable to gauge choices and an unphysical early-time instability.

ii) The coupled scalar field conformally interacting dark energy CIDE model   \cite{bekenstein1993relation,koivisto2014dark,de2015disformal,gomez2023conformally,varela2026tale}, which is derived from the Lagrangian formalism of the action. This is the case that will be considered in the current manuscript. It consists of a class of scalar-tensor theories and a metric-dependent coupling encompassing conformal, and possibly also disformal, transformations that derive the dark sector interaction from a covariant action principle where the DM particles are minimally coupled to the effective frame metric $\tilde{g}_{\mu\nu} = C(\phi)g_{\mu\nu} + D(\phi)\partial_\mu \phi \partial_\nu \phi$.  The conformal factor represented by $C(\phi)$ mediates a field-dependent fifth force that rescales the particle mass, and the disformal factor represented by $D(\phi)$ introduces a coupling that modifies momentum exchange and kinetic dynamics. This action-derivative framework ensures self-consistency in the evolution of gauge-invariant perturbations across both the linear and non-linear regimes. The interaction through conformal and disformal transformations is inspired by fundamental physics instead of being a purely phenomenological modification of $\Lambda$CDM \cite{bekenstein1993relation,zumalacarregui2010disformal}.

Historically, the action-derived CIDE model was built on the Einstein frame metric transformations pioneered by Dicke in 1962 \cite{Dicke:1962}. As for the scalar-tensor field interactions, they were originally proposed to be applied to particle physics and DM dynamics \cite{Damour:1990} and adopted in the formalism for quintessence scenarios, establishing the conformally coupled dark sectors where the metric is written as $\tilde{g}_{\mu\nu} = C(\phi)g_{\mu\nu}$  \cite{Amendola:2000,bekenstein1993relation,koivisto2014dark,de2015disformal,damour1994string,brax2004detecting, fujii2003scalar,amendola2010dark}. This approach makes a mathematically closed set of the background and perturbation equations, overcoming the gauge instability inherent to \texttt{ad hoc} fluid-level parameterizations \cite{Valiviita:2008iv}.  These studies have shown the potential of conformal and disformal coupling models to address different cosmological issues. For instance, \cite{amendola2000coupled} considered the impact of a conformal coupling between quintessence and DM  for the background history and the growth of cosmic structures. Structure formation and perturbations were investigated in \cite{pettorino2008coupled}. The observational signatures of DE with disformal scalar couplings were explored in \cite{zumalacarregui2010disformal},  the interactions with matter were addressed in  \cite{de2015disformal}, while the effect of such couplings on the background evolution was discussed in \cite{koivisto2008dynamics,de2015disformal,van2018searching}. On the other hand, linear cosmological perturbations and structure growth were studied in \cite{de2015disformal,koivisto2012clustering,pettorino2008coupled}. Within the context of inflationary cosmology, conformal and disformal couplings were discussed in \cite{bettoni2013disformal}, where the parameters of conformal and disformal interacting DE models, including the coupling strength, scalar potential, and conformal or disformal coupling coefficients were constrained using cosmological data sets such as Planck CMB, Pantheon Supernovae, BAO, and Cosmic Chronometers \cite{de2015disformal,mifsud2017probing,aghanim2020planck}.

Recently, Alvarez et al. \cite{alvarez2022eternal} constructed a framework that involves an interacting DE-DM model with conformal and disformal couplings but assuming a pressureless scalar field $P_{\phi}=0$. The scalar field, therefore, behaves as a matter-like fluid rather than as a proper kind of DE, with the model being unable to provide a component that could account for the observed late-time acceleration of the universe. That work focused on the background evolution, confronting it with the late-time cosmological dataset by assuming a power-law parametrization of the conformal function $C(\phi)= C_0 (1+\phi)^m$ and setting $D(\phi) = 0$. The main aim of the present manuscript is to provide an extension of the work by Alvarez et al. \cite{alvarez2022eternal}  by taking into account the fact that the scalar field behaves as a DE with pressure $P_{\phi}=w_{\phi}\rho_{\phi}$, where the nature of the DE will be set up by the following equation of state parameter of the canonical scalar field:
\begin{eqnarray} \label{EQ}
    w(\phi) = \frac{\dot{\phi}/2-V(\phi)}{\dot{\phi}/2+V(\phi)}\;.
\end{eqnarray}

 We also confront the model with CMB measurements from  Planck 2018 \cite{aghanim2020planck,aghanim2020planckx}. We will also use BAO data from the Dark Energy Survey (DESI DR2) \cite{andrade2025validation, abdul2025desi} and Supernova Type Ia distance compilations (PantheonPlus, PantheonPlus + SH0ES, Union3, and DES-Dovekie) \cite{2025ApJ...986..231R,brout2022pantheon+,popovic2026dark} for the \texttt{MCMC}. We shall subsequently investigate the thermodynamic properties of cosmological perturbations including CMB anisotropies, the scale-dependent transfer functions, the growth rate, and the matter power spectrum, to test how CIDE influences structure growth and whether it can mitigate the Hubble and matter clustering tensions when confronted with early and late-time observations.

The content of this work is organized as follows: In Section \ref{section 2}, the action of scalar-tensor in the Einstein frame is introduced and related to the fundamental equations of the models for interacting DE and DM, which are described through conformal and disformal couplings. In Section \ref{section 3}, we impose constraints on the model by confronting it with datasets, highlighting the cosmological tensions. The detailed statistical analysis, the background evolution of the cosmic history and the numerical diagram of the power spectrum for CMB anisotropies, the scale-dependent transfer functions, the growth factor, growth rate, and matter power spectrum from the \texttt{CLASS} code are presented to demonstrate the deviation of the interacting model from $\Lambda$CDM that comes from the exchange of energy through interactions. Finally, Section \ref{section 4} presents the conclusion of our work.

\section{Basic equations}\label{section 2}

Models for interacting DE and DM with conformal and disformal couplings can be related to the action of scalar-tensor theory in the Einstein frame as  \cite{van2015disformal,mifsud2017probing,alvarez2022eternal,xiao2019can,van2018searching}
\begin{eqnarray}\label{action}
S &=& \int d^4x \, \sqrt{-g} \bigg[ \frac{M_{\text{Pl}}^2}{2} R - \frac{1}{2} g^{\mu\nu} \partial_\mu \phi \, \partial_\nu \phi - V(\phi) + \mathcal{L}_{\text{SM}} \bigg] \nonumber\\&&+ \int d^4x \, \sqrt{-\tilde{g}} \, \tilde{\mathcal{L}}_{\text{DM}} (\tilde{g}_{\mu\nu}, \psi)\ ,
\end{eqnarray}
where \( M_{\text{Pl}} = 2.4 \times 10^{18} \) GeV is the reduced Planck mass. The uncoupled Standard Model (SM) particles are represented by the Lagrangian \( \mathcal{L}_{\text{SM}} \), which includes radiation (\( \text{r} \)) and baryonic matter (\( \text{b} \)). \( \tilde{\mathcal{L}}_{\text{DM}} \) denotes the Lagrangian for the DM fields \( \tilde{\psi}_m \). The scalar field \( \phi \), along with its kinetic and potential  {$V(\phi)$} terms, plays the role of DE that is responsible for the accelerating expansion of the universe. In this model, baryonic matter, radiation, and the scalar field (DE) components are all minimally coupled to the metric $g_{\mu\nu}$, while the DM sector is coupled to a different metric, $\tilde{g}_{\mu\nu}$.  The field equations for the action (\ref{action}) are found by variation with respect to $g_{\mu\nu}$, and can be written as
\begin{eqnarray}
    R_{\mu\nu}-\frac{1}{2}g_{\mu\nu}R = \kappa^2\left(T^{SM}_{\mu\nu}+T^{\phi}_{\mu\nu} + T^{DM}_{\mu\nu}\ \right)\;,
\end{eqnarray}
where  $\kappa^2 =  8\pi G = \frac{1}{M_{\text{Pl}}^2}$\footnote{We assume $8\pi G = \frac{1}{M_{\text{Pl}}^2} = 1$} the associated energy-momentum tensors for each collection of matter fields are defined as follows
\begin{eqnarray}
   && T^{SM}_{\mu\nu} = \frac{-2}{\sqrt{-g}}\frac{\delta \left(\sqrt{-g}\mathcal{L}_{SM}\right)}{\delta g^{\mu\nu}} \;, \\&&
   T_{\mu\nu}^{\phi}=\partial_{\mu}\phi\partial_{\nu}\phi-g_{\mu\nu}\left(\frac{1}{2}\partial_{\sigma}\phi\partial^{\sigma}\phi+V(\phi)\right)\ , \\&&
T_{\mu\nu}^{DM}=\frac{-2}{\sqrt{-g}}\frac{\delta \left(\sqrt{-\tilde{g}}\tilde{\mathcal{L}}_{DM}\right)}{\delta g^{\mu\nu}}\ ,
\end{eqnarray}
for the SM particles, DE, and DM, respectively. From the action presented in Eq. \eqref{action}, the DM fields \(\tilde{\mathcal{L}}_{\text{DM}} (\tilde{g}_{\mu\nu}, \psi)\) follow the geodesics  of the metric $\tilde{g}_{\mu\nu}$, defined as
\begin{equation}\label{geo  DESI DR2c}
    \tilde{g}_{\mu\nu}(\phi) = C(\phi) g_{\mu\nu} + D(\phi) \, \partial_\mu \phi \, \partial_\nu \phi \;,
\end{equation}
with \( C(\phi) \) and \( D(\phi) \) being the conformal and disformal coupling functions, respectively. This way, DM particles' trajectories are determined by both the metric $g_{\mu\nu}$, the DE component, and the conformal and disformal functions.

As a consequence of the interaction between DM and DE, neither the energy-momentum tensor of the scalar field nor that of DM are conserved and, instead, they satisfy the equations
\[
 \quad \nabla^\mu T^{(DM)}_{\mu\nu} =  Q \nabla_\nu\phi\;,\quad \text{and} \quad \nabla^\mu T^{(\phi)}_{\mu\nu} = -Q \nabla_\nu\phi\;.
\]
However, the total energy-momentum tensor is conserved for all fluids combined, that is $$\nabla^\mu T^{(total)}_{\mu\nu} \equiv \nabla^\mu \left( T^{(SM)}_{\mu\nu} + T^{(DM)}_{\mu\nu} + T^{(\phi)}_{\mu\nu}\right) = 0.$$  
The corresponding conservation equations for each fluid can be written as:
\begin{eqnarray}
&&\dot{\rho}_{r}+4 H \rho_{r}=0\;,\\&& \dot{\rho}_{b}+3H \rho_{b}=0\;,\\ &&  \dot{\rho}_{DM}+3H \rho_{DM}=-Q\dot{\phi}\;,\label{continoutydarkmatter}\\&&
\ddot{\phi}+3H\dot{\phi}+\frac{dV(\phi)}{d\phi}=Q \label{continoutydarkenergy}\;,
\label{conteqs2}
\end{eqnarray}
where $\rho_{r}$  is the energy density of the radiation,  $\rho_b$ is the baryonic density,  $\rho_{DM} $ is the DM density, and $\rho_{\phi}$ is the DE density. The energy density and pressure of the scalar field can be written as
\be
\rho_{\phi}=\frac{1}{2}\dot{\phi}^2+V(\phi)\ , \quad p_{\phi}=\frac{1}{2}\dot{\phi}^2-V(\phi)\ .
\label{phiRhoP}
\ee
Its equations of motion are found through variation with respect to the scalar field $\phi$, which provides the equation \cite{van2017testing}
\begin{eqnarray}\label{scalrfieldequation}
    \Box\phi-\frac{dV(\phi)}{d\phi}=-Q\ ,
\end{eqnarray}
and the equation of state parameter of the canonical scalar field becomes
\begin{eqnarray} \label{EQ}
    w(\phi) = \frac{\dot{\phi}/2-V(\phi)}{\dot{\phi}/2+V(\phi)} \;.
\end{eqnarray}
 Consequently,  the continuity equations for DM and DE presented in Eqs.\eqref{continoutydarkmatter} and \eqref{continoutydarkenergy} read as
\begin{eqnarray}
    &&\dot{\rho}_{DM}+3H \rho_{DM}=-{Q}\dot{\phi}\ .\nonumber\\&&
    \dot{\rho}_{\phi}+3H \rho_{\phi}(1+w_{\phi})={Q}\dot{\phi}\ .
\end{eqnarray}
Following the approach of \cite{alvarez2022eternal,xiao2019can, van2018searching}, the flow of energy and momentum between the two dark fluids (DM and DE) is governed by the interaction term  $Q$, which can be expressed in terms of the conformal and disformal functions as  
\begin{eqnarray}\label{intractionterm}
   Q=&&\frac{C'(\phi)}{2C(\phi)}g^{\mu\nu}T^{DM}_{\mu\nu}+\frac{D'(\phi)}{2C(\phi)}T^{DM\mu\nu}\partial_{\mu}\phi\partial_{\nu}\phi\nonumber\\&& - \nabla_{\mu}\left[\frac{D(\phi)}{C(\phi)}T_{DM}^{\mu\nu}\nabla_{\nu}\phi\right]\ . 
\end{eqnarray}

For the sake of comparison with \cite{alvarez2022eternal}, here we will stick to a purely conformal coupling, neglecting potential disformal interactions, $D(\phi)=0$.  In this reduced scenario, the interacting term \eqref{intractionterm} boils down to
\be
{Q}=\frac{C_\phi (\phi)}{2C(\phi)}g^{\mu\nu}{T}_{\mu\nu}^{DM}=- \frac{C_\phi (\phi)}{2C(\phi)}\rho_{DM}\ .
\label{QConf}
\ee
We will assume a power-law ansatz for the conformal function as
\begin{eqnarray}
    C(\phi) = C_0 (1+\phi)^m \;. \label{conservat}
\end{eqnarray}
 Consequently, the coupled background equations are represented by
 \begin{eqnarray}
     && \rho'_{DM}  +3\mathcal{H}\rho_{DM} = \zeta(\phi)\phi' \rho_{DM} \label{DM} \;, \\&&
     \phi '' +2\mathcal{H}\phi'  = - a^2 \left(\frac{dV}{d\phi} +3 \zeta(\phi)\rho_{DM}\right) \;, \label{PHI}
 \end{eqnarray}
where $$\zeta(\phi) = \frac{m}{2(1+\phi)} \;, $$ prime prime  $''$ is $d^2/d\tau^2$, prime  $'$ is $d/d\tau$ and the conformal Hubble parameter $\mathcal{H} \equiv aH$. 
Then, from  Eq. \eqref{PHI}, the equivalent  scalar-field energy balance is  
\begin{eqnarray}
\dot{\rho}_{\phi}+3H (\rho_{\phi}+P_{\phi})= \zeta(\phi) \phi' \rho_{DM}\;. \label{finalcons1}
\end{eqnarray}
The pressureless condition $P_{\phi}=0$ adopted in Ref. \cite{alvarez2022eternal}, indicates that their scalar field behaves as a matter-like fluid rather than as DE. Since that choice does not provide a component able to account for the observed late-time cosmic acceleration, here we relax that condition and consider that the scalar field behaves as a DE fluid with $P_{\phi}=w_{\phi}\rho_{\phi}$, where $w(\phi)$ is presented in Eq. \eqref{EQ}. Note that for $m=0$ and for a constant scalar field ($\dot{\phi}=0$), the standard model of cosmology is recovered. For $m = 0$, the uncoupled scalar-field DE is recovered and   the standard matter scenario holds: there is no particle creation or interaction.  For $m>0$, DM loses energy through decay, annihilation, or transfers energy to another component, which in our case is DE. For $m<0$, the model describes DM creation or energy transfer from another source, such as DE. 

We now fully shift our attention to the perturbation level. We choose to work in the conformal-Newtonian gauge perturbed FLRW metric as \cite{malik2009cosmological,mukhanov1992theory,1995ApJ...455....7M}
\begin{equation}
    ds^{2}=a^{2}(\tau)\Big[-(1+2\Psi)d\tau^{2}+(1-2\Phi)\delta^{ij}dx_{i}%
dx_{i}\Big]\;,
\end{equation}
where $\Phi$ represents the Newtonian gravitational potential, and $\Psi$ corresponds to the perturbations of the spatial curvature.  For the  perturbed universe, the energy-momentum tensor  for the two dark fluids reads as \cite{SINHA2023101273,cheng2020testing,li2014large}
\begin{align}
\delta\left(  \nabla_{\mu}T_{DE}^{\mu\nu}\right)  =-\delta\left(
\nabla_{\mu}\delta T_{DM}^{\mu\nu}\right)  =\delta \mathcal{Q}_{i}^{\nu}
\;. \label{diffusivex}%
\end{align}
From the conservation equations \eqref{DM}, and  \eqref{PHI}, it is evident that the conformal framework yields a covariant construction in which DM evolves along a metric conformally coupled to the spacetime geometry via scalar field-dependent functions.

For the  present model the interaction term is scalar-dependent, $\mathcal{Q}^\mu \propto {Q}\nabla^\mu\phi$, and the linear scalar perturbation becomes $\phi(t,x)=\bar{\phi}(t)+\delta\phi(t,x)$, with the corresponding perturbed interaction being 
\begin{eqnarray*}
    &&\delta\mathcal{Q}^\mu= \frac{{1}}{a^2} \frac{C_\phi (\phi)}{2C(\phi)}\rho_{DM}\bigg(     \left[\delta\phi'+ {\phi}'\left(\delta_{DM}-a\delta\phi-2\Phi\right)\right],\partial^i\delta\phi\bigg)\;,
\end{eqnarray*}
where $\delta \phi$ is the  scalar-field perturbations. Eq. \eqref{diffusivex} represents the generalized form of the conservation equation of the density contrast, which governs the density perturbations. The Newtonian gauge's conservation equations for the density contrast $\delta_{DM}$ and momentum $\theta_{DM} = ik_jv^j_{DM}$ of the dark matter fluid are given by \cite{valiviita2008large,majerotto2010adiabatic}
\begin{align}\label{theta}
  &{\delta'_{DM}}  =  3{\Phi'}-\theta_{DM}  + \zeta(\phi) \delta \phi' + \frac{d\zeta(\phi)}{d\phi} \phi'\delta \phi \;,\\
&  {\theta'_{DM}}  -\left(\mathcal{H+\zeta(\phi)}\right)\theta_{DM} +\kappa^2 \Psi _\kappa^2\zeta(\phi) \delta \phi\;. \label{delta}
\end{align}
From the above two key conservation equations \eqref{theta} and \eqref{delta},  one can show that the density contrast eventually grows for each fluid, and $\theta_i$ shows how the velocity perturbations are pushed due to the pressure of the fluid and the gravitational potential $\Phi$ and $\Psi$.
The perturbed Klein-Gordon equation is 
\begin{eqnarray} \label{XX}
       \delta \phi'' 
    =& 4\phi' \Phi' -2a^2 \left(\frac{dV}{d\phi}+3\zeta(\phi)\rho_{DM}\right)\Psi \nonumber\\&-3a^2 \zeta(\phi)\rho_{DM}\delta_{DM} \;, 
\end{eqnarray}
and the scalar-field momentum density equation becomes,
\begin{eqnarray}
    (\rho_\phi +p_\phi)\theta_\phi = \frac{1}{3}\frac{k^2}{a^2}\phi' \delta \phi\;.
\end{eqnarray}

The  above Eqs. \eqref{theta}  and \eqref{XX} are the key coupled differential equations for further analysis of the perturbative dynamics, and these equations are reduced to the uncoupled scalar-field DE (UCSFDE) system when
$m =0$. In most cases, the cosmological instability for the phenomenological model is unavoidable when the equation of state parameter for dark energy is $w = -1$. However, such work considers that the action-driven scalar-field dark energy mode is free from the cosmological instability, unlike the phenomenological model at the perturbation level; see Eq. \eqref{XX}.  The Einstein-Boltzmann system of equations is commonly computed using the polynomial-featured Cosmic Linear Anisotropy Solving System \texttt{CLASS} code \cite{lesgourgues2011cosmic} to constrain observable parameters, including the extra term from the  CIDE model.  During the implementation of the modified \texttt{CLASS} code, we consider the Albrecht-Skordis potential as Ref. \cite{Albrecht:1999rm}
\begin{eqnarray}
      V(\phi) = e^{-\lambda\phi/M_{pl}}\left(A-B)^\alpha + A \right);,
\end{eqnarray}
  where reduced Planck  mass $M_p$, and the other parameters $\lambda$, $\alpha$, $A$ and  $B$ are the scalar field parameters. For more simplicity, we assume the default \texttt{CLASS} values of  $ \alpha = A = B = 0$ to reduce to the polynomial-feature scalar potential to a pure  simple exponential potential as $V(\phi) = e^{-\lambda \phi} $, see more Ref. \cite{lesgourgues2011cosmic,ferreira1998cosmology}. 

In the upcoming section, we confront our model of the background cosmological evolution together with the perturbative thermodynamic evolution, using equations  \ref{DM}, \ref{PHI}, \eqref{theta} \eqref{delta} and \eqref{XX} accordingly, using the modified CLASS in the Newtonian gauge, and constraining the parameters using \texttt{COBAYA}. We strictly adopted the scalar field potential implementations in the CLASS code using the exponential Albrecht-Skordis potential with the field expression in Planck mass units $M_{p} =1$. We specified that the initial conditions of $\phi_{\rm in} = 100$ and $\phi'_{\rm in} = 0$ without imposing the attractor initial conditions. In this case, the scalar field acts as a dark energy density, while the contribution of the dark energy fluid and the cosmological constant is set to zero in our \texttt{.yaml} file, see morw in Ref.  \cite{lesgourgues2011cosmic}. 

At each sampled cosmological parameter point, CLASS determines $\lambda$ through its shooting procedure to satisfy $\Omega_{\phi} = 1- \sum_i \Omega_i$, where $i$ represents the species of the fluid that is considered. This practical implementation helps us to constrain our CIDE scenario and show the influence of the flow of energy by the coupling parameter $m$, and its effect on the structure growth of the universe.

\section{Results and Discussion}\label{section 3}

We shall now proceed to confront our CIDE models with observational data. Our results are organized as follows: 

\begin{enumerate}[label=\roman*)]

\item First, we carry out \texttt{MCMC} simulations to constrain the models with recent cosmological data. For this purpose, we perform a Bayesian analysis employing the \texttt{COBAYA}\footnote{cobaya.readthedocs.io/en/latest/installation.html} interface \cite{torrado2021cobaya}, where the cosmological equations are solved using the modified \texttt{CLASS}\footnote{class-code.net/} code \cite{lesgourgues2011cosmic}. For the analysis of the numerical chains, we make use of \texttt{GetDist}\footnote{https://getdist.readthedocs.io/en/latest/} to obtain the cosmological parameter constraints. We then apply the Akaike Information Criterion (AIC) and the Bayesian/Schwarz Information Criterion (BIC) to classify the models according to their compatibility with the observational data.

\item Second, we highlight the discrepancy between our model's inferred value of $H_0$ and both the early-time measurement from Planck 2018 ($67.4 \pm 0.5~\mathrm{km\,s^{-1}\,Mpc^{-1}}$) \cite{aghanim2020planck} and the late-time SH0ES 2022 measurement ($73.04 \pm 1.04~\mathrm{km\,s^{-1}\,Mpc^{-1}}$) \cite{Riess:2021jrx,riess2019large}, demonstrating our model's potential to alleviate the Hubble tension.

\item Finally, we examine the perturbative thermodynamic properties of the model in comparison with standard $\Lambda$CDM cosmology, and we evaluate the transfer function for matter density contrast $T_m(z,k)$, the growth rate $f(z)$, the growth factor $D(z)$, the redshift-space distortion parameter $f\sigma_8(z)$, the matter power spectrum $P_m(k,z)$, and the CMB anisotropy power spectrum to trace the evolution of the Universe and the growth of cosmic structures.
\end{enumerate}

\begin{table*}[htbp]
\centering
\renewcommand{\arraystretch}{1.3}
\caption{Marginalized $\Lambda$CDM, UCSFDE and CIDE constraints at 95\% confidence. $H_0$ is in km\,s$^{-1}$\,Mpc$^{-1}$, $t_0$ in Gyr, and $r_\mathrm{drag}$ in Mpc.}
\label{tab:constraints}
\scriptsize
\begin{tabular}{lcccccc}
\hline
\hline
Parameter & \texttt{CMB} & \texttt{CMB +   DESI  DR2} & \texttt{CMB +   DESI  DR2} & \texttt{CMB +   DESI  DR2}  & \texttt{CMB +   DESI  DR2} & \texttt{CMB +   DESI  DR2}  \\
& & &  \texttt{+  DD} &   \texttt{+ PP} & \texttt{ + PPS} & \texttt{+U3}\\
\hline
\hline
\multicolumn{7}{l}{\textbf{$\Lambda$CDM}} \\
\hline
$\ln(10^{10}A_\mathrm{s})$ & $3.0456^{+0.0221}_{-0.0237}$ & $3.0540^{+0.0316}_{-0.0273}$ & $3.0498^{+0.0335}_{-0.0290}$ & $3.0524^{+0.0306}_{-0.0340}$ & $3.0546^{+0.0326}_{-0.0313}$ & $3.0517^{+0.0311}_{-0.0294}$ \\
$n_\mathrm{s}$ & $0.96460^{+0.00723}_{-0.00696}$ & $0.97016^{+0.00733}_{-0.00743}$ & $0.96974^{+0.00660}_{-0.00706}$ & $0.97020^{+0.00712}_{-0.00691}$ & $0.97171^{+0.00655}_{-0.00575}$ & $0.96933^{+0.00696}_{-0.00729}$ \\
$H_0$ & $67.379^{+0.974}_{-0.996}$ & $68.408^{+0.621}_{-0.651}$ & $68.287^{+0.548}_{-0.524}$ & $68.324^{+0.582}_{-0.595}$ & $68.754^{+0.625}_{-0.595}$ & $68.251^{+0.611}_{-0.584}$ \\
$\Omega_\mathrm{b}h^2$ & $0.022379^{+0.000272}_{-0.000263}$ & $0.022549^{+0.000252}_{-0.000253}$ & $0.022510^{+0.000243}_{-0.000290}$ & $0.022483^{+0.000265}_{-0.000262}$ & $0.022634^{+0.000282}_{-0.000255}$ & $0.022513^{+0.000269}_{-0.000242}$ \\
$\Omega_\mathrm{c}h^2$ & $0.11998^{+0.00222}_{-0.00219}$ & $0.11773^{+0.00162}_{-0.00154}$ & $0.11798^{+0.00140}_{-0.00142}$ & $0.11777^{+0.00114}_{-0.00112}$ & $0.11706^{+0.00132}_{-0.00125}$ & $0.11807^{+0.00138}_{-0.00150}$ \\
$\Omega_\mathrm{m}$ & $0.3151^{+0.0134}_{-0.0123}$ & $0.30117^{+0.00881}_{-0.00838}$ & $0.30268^{+0.00730}_{-0.00730}$ & $0.30185^{+0.00705}_{-0.00683}$ & $0.29690^{+0.00820}_{-0.00732}$ & $0.30320^{+0.00757}_{-0.00846}$ \\
$S_8$ & $0.8312^{+0.0237}_{-0.0229}$ & $0.8096^{+0.0194}_{-0.0183}$ & $0.8109^{+0.0139}_{-0.0149}$ & $0.8103^{+0.0196}_{-0.0179}$ & $0.8020^{+0.0143}_{-0.0152}$ & $0.8125^{+0.0181}_{-0.0194}$ \\
$t_0$ & $13.7961^{+0.0416}_{-0.0420}$ & $13.7610^{+0.0334}_{-0.0310}$ & $13.7659^{+0.0374}_{-0.0322}$ & $13.7689^{+0.0357}_{-0.0354}$ & $13.7463^{+0.0365}_{-0.0387}$ & $13.7666^{+0.0326}_{-0.0365}$ \\
$r_\mathrm{drag}$ & $147.095^{+0.463}_{-0.445}$ & $147.504^{+0.473}_{-0.508}$ & $147.480^{+0.461}_{-0.486}$ & $147.565^{+0.365}_{-0.360}$ & $147.589^{+0.419}_{-0.414}$ & $147.453^{+0.416}_{-0.433}$ \\
\hline
\multicolumn{7}{l}{\textbf{UCSFDE}} \\
\hline
$\ln(10^{10}A_\mathrm{s})$ & $3.0458^{+0.0311}_{-0.0280}$ & $3.0545^{+0.0292}_{-0.0312}$ & $3.0529^{+0.0310}_{-0.0328}$ & $3.0500^{+0.0246}_{-0.0237}$ & $3.0583^{+0.0310}_{-0.0280}$ & $3.0584^{+0.0347}_{-0.0315}$ \\
$n_\mathrm{s}$ & $0.96383^{+0.00797}_{-0.00787}$ & $0.97020^{+0.00728}_{-0.00674}$ & $0.96934^{+0.00626}_{-0.00627}$ & $0.96956^{+0.00617}_{-0.00682}$ & $0.97124^{+0.00622}_{-0.00721}$ & $0.96977^{+0.00631}_{-0.00619}$ \\
$H_0$ & $67.230^{+1.100}_{-1.220}$ & $68.326^{+0.565}_{-0.609}$ & $68.197^{+0.537}_{-0.528}$ & $68.232^{+0.544}_{-0.498}$ & $68.674^{+0.574}_{-0.526}$ & $68.233^{+0.641}_{-0.551}$ \\
$\Omega_\mathrm{b}h^2$ & $0.022353^{+0.000303}_{-0.000315}$ & $0.022524^{+0.000234}_{-0.000244}$ & $0.022474^{+0.000230}_{-0.000223}$ & $0.022496^{+0.000226}_{-0.000229}$ & $0.022620^{+0.000238}_{-0.000221}$ & $0.022491^{+0.000257}_{-0.000233}$ \\
$\Omega_\mathrm{c}h^2$ & $0.12016^{+0.00269}_{-0.00235}$ & $0.11772^{+0.00126}_{-0.00122}$ & $0.11793^{+0.00138}_{-0.00126}$ & $0.11790^{+0.00130}_{-0.00115}$ & $0.11707^{+0.00125}_{-0.00131}$ & $0.11788^{+0.00130}_{-0.00140}$ \\
$\Omega_\mathrm{m}$ & $0.3169^{+0.0157}_{-0.0151}$ & $0.30182^{+0.00751}_{-0.00718}$ & $0.30331^{+0.00739}_{-0.00701}$ & $0.30298^{+0.00663}_{-0.00670}$ & $0.29759^{+0.00690}_{-0.00697}$ & $0.30291^{+0.00728}_{-0.00783}$ \\
$S_8$ & $0.8337^{+0.0269}_{-0.0249}$ & $0.8105^{+0.0170}_{-0.0163}$ & $0.8125^{+0.0182}_{-0.0169}$ & $0.8108^{+0.0146}_{-0.0152}$ & $0.8040^{+0.0171}_{-0.0169}$ & $0.8141^{+0.0183}_{-0.0169}$ \\
$t_0$ & $13.8002^{+0.0492}_{-0.0507}$ & $13.7637^{+0.0324}_{-0.0327}$ & $13.7708^{+0.0312}_{-0.0328}$ & $13.7678^{+0.0306}_{-0.0325}$ & $13.7479^{+0.0312}_{-0.0313}$ & $13.7686^{+0.0326}_{-0.0327}$ \\
$r_\mathrm{drag}$ & $147.074^{+0.488}_{-0.493}$ & $147.533^{+0.359}_{-0.375}$ & $147.532^{+0.415}_{-0.427}$ & $147.516^{+0.345}_{-0.395}$ & $147.601^{+0.383}_{-0.387}$ & $147.528^{+0.408}_{-0.393}$ \\
\hline
\multicolumn{7}{l}{\textbf{CIDE}} \\
\hline
$\ln(10^{10}A_\mathrm{s})$ & $3.0415^{+0.0296}_{-0.0276}$ & $3.0556^{+0.0273}_{-0.0270}$ & $3.0539^{+0.0281}_{-0.0261}$ & $3.0498^{+0.0281}_{-0.0282}$ & $3.0527^{+0.0277}_{-0.0287}$ & $3.0517^{+0.0269}_{-0.0277}$ \\
$n_\mathrm{s}$ & $0.96389^{+0.00972}_{-0.00979}$ & $0.96619^{+0.00764}_{-0.00751}$ & $0.96594^{+0.00677}_{-0.00701}$ & $0.96444^{+0.00766}_{-0.00735}$ & $0.96403^{+0.00734}_{-0.00782}$ & $0.96651^{+0.00851}_{-0.00919}$ \\
$H_0$ & $67.760^{+4.170}_{-3.870}$ & $68.908^{+0.955}_{-0.947}$ & $68.655^{+0.756}_{-0.856}$ & $68.773^{+0.854}_{-0.831}$ & $69.528^{+0.789}_{-0.833}$ & $68.747^{+0.891}_{-0.959}$ \\
$\Omega_\mathrm{b}h^2$ & $0.022368^{+0.000553}_{-0.000366}$ & $0.022372^{+0.000282}_{-0.000341}$ & $0.022354^{+0.000310}_{-0.000295}$ & $0.022314^{+0.000293}_{-0.000284}$ & $0.022341^{+0.000295}_{-0.000328}$ & $0.022347^{+0.000347}_{-0.000319}$ \\
$\Omega_\mathrm{c}h^2$ & $0.11939^{+0.00501}_{-0.00520}$ & $0.11757^{+0.00111}_{-0.00114}$ & $0.11790^{+0.00131}_{-0.00132}$ & $0.11781^{+0.00127}_{-0.00131}$ & $0.11709^{+0.00135}_{-0.00139}$ & $0.11775^{+0.00129}_{-0.00130}$ \\
$m$ & $0.0012^{+0.014}_{-0.014}$ & $0.0054^{+0.006}_{-0.006}$ & $0.0049^{+0.004}_{-0.005}$ & $0.0065^{+0.005}_{-0.005}$ & $0.008^{+0.005}_{-0.005}$ & $0.004^{+0.006}_{-0.006}$ \\
$\Omega_\mathrm{m}$ & $0.3114^{+0.0511}_{-0.0521}$ & $0.29612^{+0.00985}_{-0.00942}$ & $0.29896^{+0.00993}_{-0.00854}$ & $0.29767^{+0.00926}_{-0.00962}$ & $0.28980^{+0.00930}_{-0.00812}$ & $0.29785^{+0.01065}_{-0.00973}$ \\
$S_8$ & $0.8263^{+0.0386}_{-0.0453}$ & $0.8199^{+0.0187}_{-0.0176}$ & $0.8212^{+0.0173}_{-0.0182}$ & $0.8204^{+0.0180}_{-0.0183}$ & $0.8198^{+0.0200}_{-0.0208}$ & $0.8196^{+0.0199}_{-0.0191}$ \\
$t_0$ & $13.772^{+0.230}_{-0.240}$ & $13.7113^{+0.0757}_{-0.0729}$ & $13.7268^{+0.0632}_{-0.0578}$ & $13.7183^{+0.0631}_{-0.0645}$ & $13.6645^{+0.0591}_{-0.0575}$ & $13.7221^{+0.0744}_{-0.0724}$ \\
$r_\mathrm{drag}$ & $147.092^{+0.562}_{-0.550}$ & $147.158^{+0.555}_{-0.590}$ & $147.188^{+0.570}_{-0.532}$ & $147.150^{+0.584}_{-0.515}$ & $146.926^{+0.526}_{-0.514}$ & $147.199^{+0.556}_{-0.623}$ \\
\hline
\hline
\end{tabular}
\end{table*}

\subsection{Parameter Constraints}

To constrain the cosmological parameters of our models, we use the following early and late-time cosmological datasets: 
\begin{itemize}
\item {CMB:} From the \textit{Planck} 2018 legacy release, we use the high-$\ell$ \texttt{Plik} TT likelihood ($30 \leq \ell \leq 2508$), TE and EE ($30 \leq \ell \leq 1996$), low-$\ell$ TT ($2 \leq \ell \leq 29$), low-$\ell$ EE ($2 \leq \ell \leq 29$), and the CMB lensing power spectrum \cite{aghanim2020planck}. We refer to this dataset as \texttt{CMB}.

 \item \texttt{BAO:}  We have used the Baryonic Acoustic Oscillation data from the DE Spectroscopic Instrument (DESI DR2) Survey released in 2024 \cite{andrade2025validation, abdul2025desi}. The measurements include 1) data for the isotropic BAO measurements of $D_V(z)/r_d$, where $D_V(z)$ and $r_d$ are the spherically averaged volume distance and sound horizon at baryon drag, respectively, 2)  anisotropic BAO measurements of $D_M(z)/r_d$ and $D_H(z)/r_d$, where $D_M(z)$ and $D_H(z)$ are the comoving angular diameter distance and the Hubble distance, respectively, and 3) the correlations between the isotropic and anisotropic BAO measurements.  Hereafter, we refer to this dataset as \texttt{DESI DR2}.

\item For the Supernova Type Ia (SNIa) dataset compilations, we have considered the SNIa distance modulus measurements from the Pantheon+ sample \cite{brout2022pantheon+}, which consists of 1701 light curves of 1550 distinct SNIa ranging in the redshift interval $z \in [0.001, 2.26]$, and we refer to this as \texttt{PP}. We also consider the Pantheon+ sample with the SH0ES Cepheid-based calibration that uses the absolute-magnitude/$H_0$ anchor, and we refer to this as \texttt{PPS}. Furthermore, we use the SN data from the Union3.0 \cite{2025ApJ...986..231R} catalogue, the most recent compilation, including 2,087 events in the redshift range $10^{-3}<z<2.27$, of which 1,363 are shared with PP, and we refer to this as \texttt{U3}.  The other recent data from the DE Survey supernova program, a reanalysis of cosmology results, and evidence for evolving DE with an updated Type Ia supernova calibration \cite{popovic2026dark} are considered from the photometric calibration of the five-year data from the DE Survey of Type Ia supernovae (DES-SN5YR). This catalogue includes 1820 SNIa events with redshifts $z<1.13$. We refer to this as \texttt{DD}.
\end{itemize}

To obtain more robust constraints on the cosmological parameters, we consider the following combinations of the early and late-time cosmological observations: 
    \begin{enumerate}
      \item \texttt{CMB}, 
    \item \texttt{CMB +   DESI DR2}, 
        \item \texttt{CMB+  DESI DR2+PP}, 
        \item \texttt{CMB+  DESI DR2+DD} and 
        \item \texttt{CMB+  DESI DR2+U3}.
    \end{enumerate}

\begin{figure*}[ht!]
    \centering
    \includegraphics[width=1.0\linewidth]{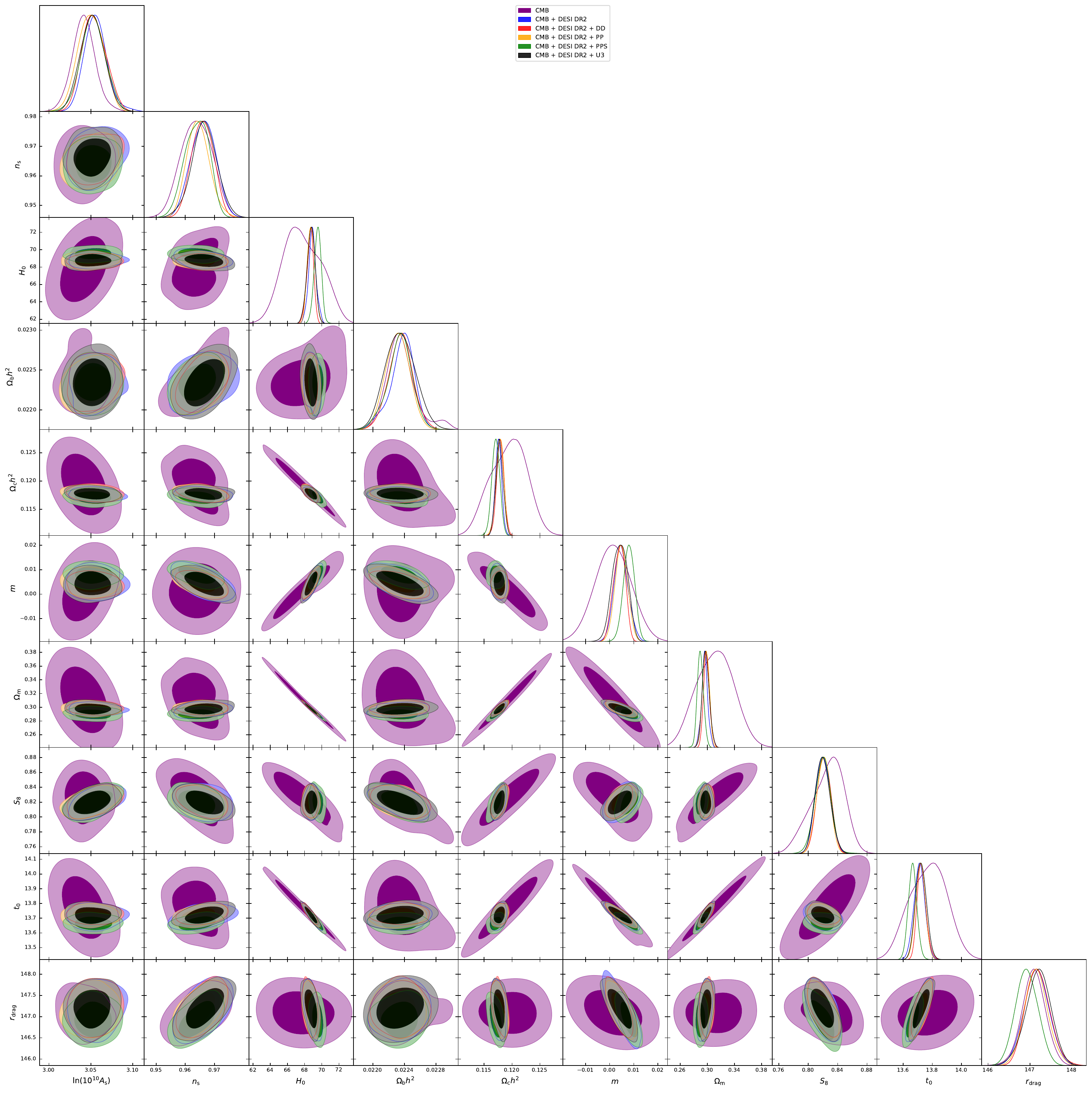}
    \caption{ The posterior distributions of the cosmological parameters for the CIDE  model using the combined early and late-time universe measurements.}
    \label{fig:placeholder1}
\end{figure*}

\begin{figure*}[ht]
    \centering
    \includegraphics[width=1.0\linewidth]{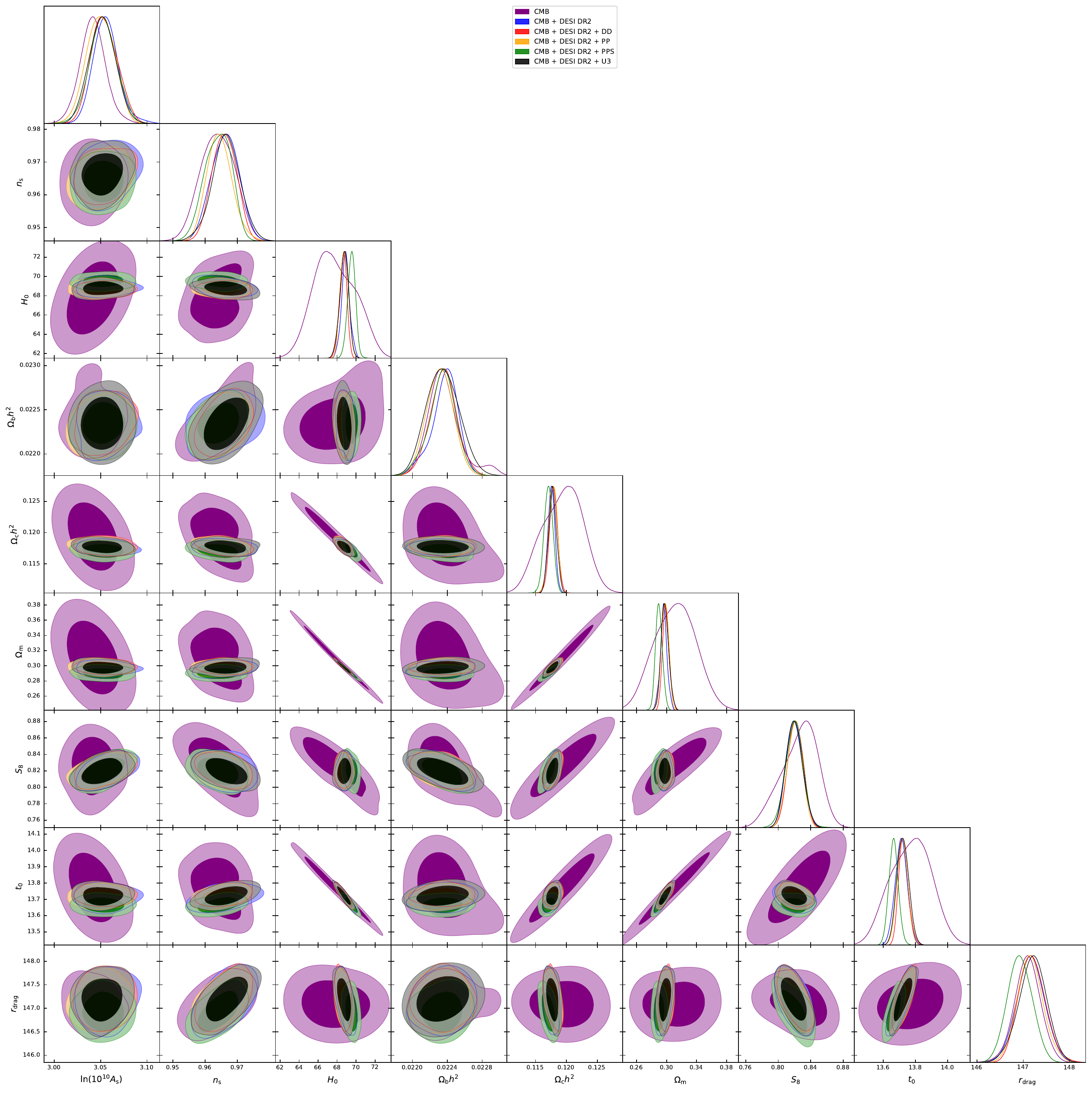}
    \caption{The posterior distributions of the cosmological parameters for the UCSFDE  model using the combined early- and late-time universe measurements.}
    \label{fig:placeholder2}
\end{figure*}

To perform parameter estimation, we adopt the following flat priors: $n_s \in [0.8, 1.2]$, $\Omega_b h^2 \in [0.0001, 0.1]$, $\Omega_{\mathrm{dm}} h^2 \in [0.001, 0.99]$, $w \in [-3.00, -0.33]$, and $m \in [-0.1, 0.1]$ for the \texttt{MCMC} simulations. The resulting marginalized constraints on these parameters are summarized in Table~\ref{tab:results_95} for the CIDE, UCSFDE, and $\Lambda$CDM models. The corresponding posterior probability distributions obtained from the MCMC analysis using the combined early and late-universe measurements are shown in Figs.~\ref{fig:placeholder1} and \ref{fig:placeholder2} for CIDE and UCSFDE,  respectively. At the 95\% confidence level (C.L.), the observational constraints favor a positive coupling parameter ($m > 0$) across all dataset combinations in both models. This consistently implies an energy transfer from DM to DE ($Q < 0$), in agreement with our theoretical predictions. 


\begin{table*}[htbp]
\centering
\renewcommand{\arraystretch}{1.3}
\caption{The statistical values of  $\chi^2$, AIC, and BIC with the corresponding deviation from the standard model of cosmology. }
\label{tab:results_95}
\small
\begin{tabular}{l|c|c|c|c|c|c}
\hline
 & \texttt{CMB} & \texttt{CMB +   DESI  DR2}  & \texttt{CMB +   DESI  DR2} & \texttt{CMB +   DESI  DR2}  & \texttt{CMB +   DESI  DR2} & \texttt{CMB +   DESI  DR2} \\
& & &  \texttt{+  DD} &   \texttt{+ PP} & \texttt{ + PPS} & \texttt{+U3}\\
 \hline
{$\chi^2_{\rm min}$}&&& & &&\\
CIDE &2777.21 & 2793.70 & 4427.11 & 4196.10 &4269.58&2821.59\\
 UCSFDE &2775.11 & 2796.38 & 4429.03 & 4197.64 &4278.24&2826.00\\
$\Lambda$CDM &2777.83& 2794.97& 4431.58 & 4204.03 &4277.19& 2824.03\\
\hline
{AIC/$\Delta$AIC}&&& &&&\\
CIDE & 2797.21/1.380& 2813.70/0.730 & 4447.11/2.47 & 4216.10 /6.230&4289.58/5.689&2841.59/0.440\\
 UCSFDE &  2793.11/2.720& 2814.38/1.410& 4447.03/2.550&4215.64/6.689& 4296.24/1.050& 2844.00/1.970\\

$\Lambda$CDM& 2795.83/------ &2812.97/------&  4449.58/------ & 4222.03/------& 4295.19/------&2842.03/------ \\
\hline
{BIC/$\Delta$BIC}&&& & &&\\
CIDE &2854.84/7.10& 2871.384/6.498 & 4510.5/ 4.419& 4278.93/0.354 &4352.575/0.614&2899.368/5.337\\
 UCSFDE  &2844.995/9.845& 2866.297/1.411 & 4504.081/2.00 & 4272.187/6.288 &4352.935/1.05&2896.590/2.559\\

$\Lambda$CDM &2847.734/------& 2864.886/------&  4506.081/------ & 4278.576/------ &4351.885/------& 2894.0302/------\\
\hline

\end{tabular}
\end{table*}
\begin{figure*}
    \centering
    \includegraphics[width=0.45\linewidth]{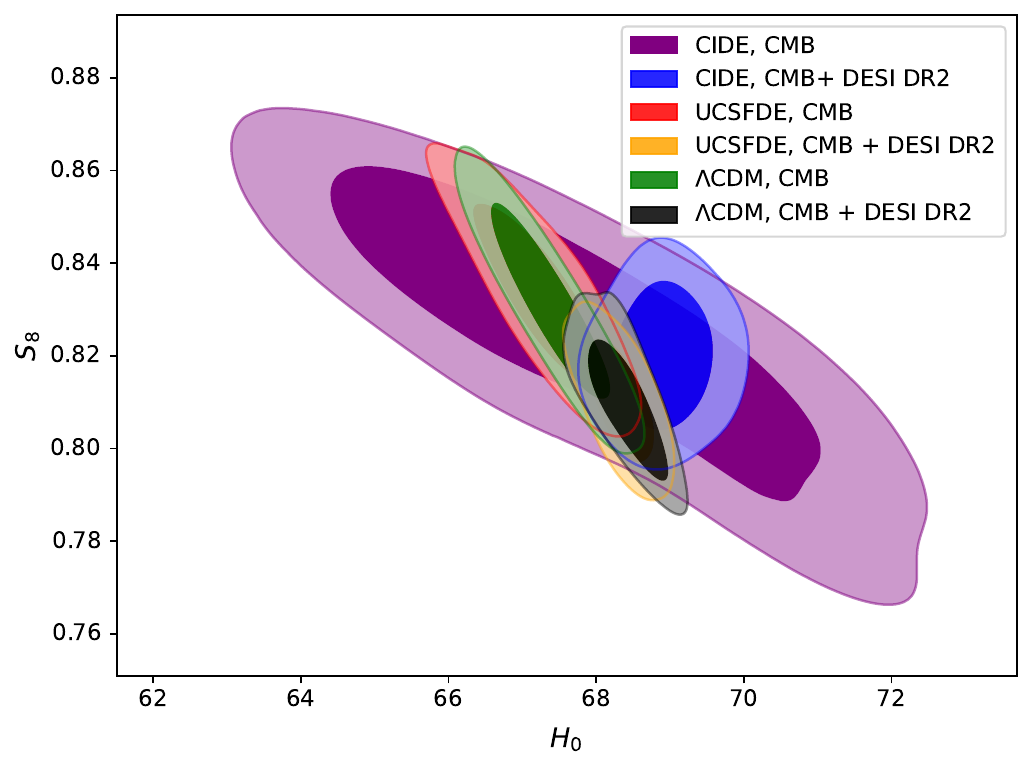}
    \includegraphics[width=0.45\linewidth]{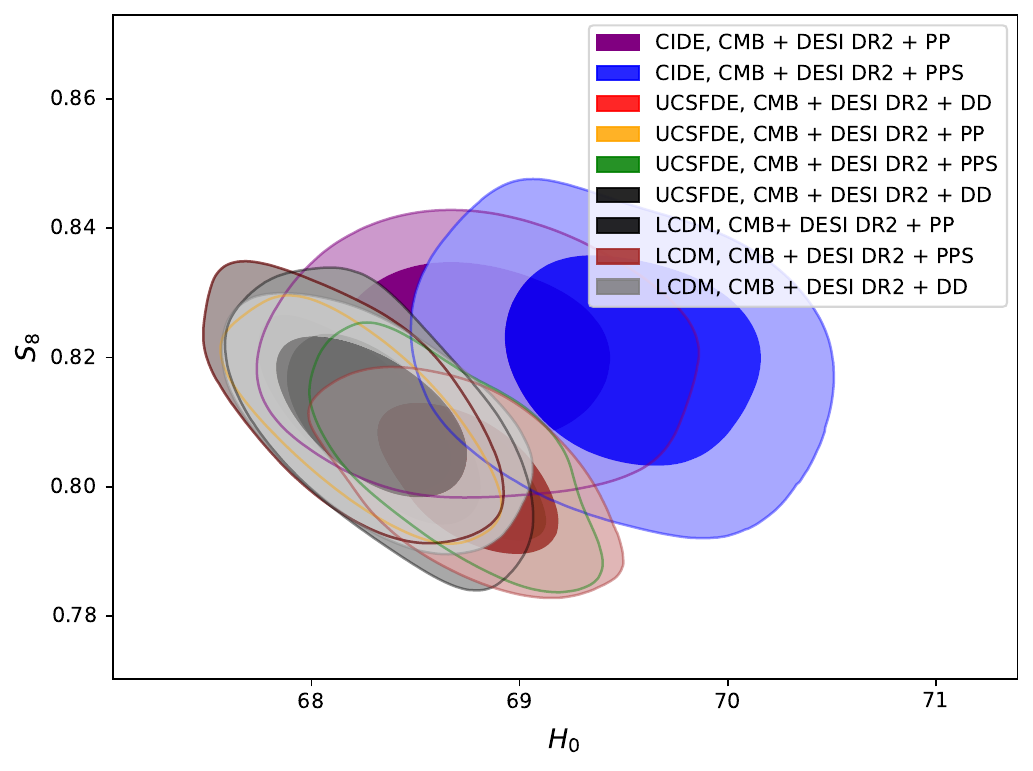}
    \caption{$S_8$ and $H_0$ constraining for CIDE, UCSFDE and $\Lambda$CDM models for different datasets. }
    \label{fig:tension1}
\end{figure*}

\subsection{Statistical analysis}
To test the viability of the model against the observational datasets, we calculate the corresponding minimum chi-squared $\chi^2$ from the likelihood of each dataset as follows: 
$$\chi^2_{\rm tottal} = \chi^2_{\rm \texttt{CMB}} + \chi^2_{\rm SN (\texttt{PP,PPS,DD,U3})} + \chi^2_{\rm {\texttt{  DESI DR2}}}\;$$
We compute the values of the Akaike Information Criterion ($\rm{AIC}=\chi^{2}+2K$) and the Bayesian/Schwarz Information Criterion ($\rm{BIC}=\chi^{2}+ K\log(N_i)$), where $K$ is the number of free parameters for that particular model and $N_i$ is the number of data points for the $i^{\rm th}$ dataset, to conduct a more thorough statistical analysis of our model in comparison with the reference model $\Lambda$CDM.
For comparison and illustrative purposes, the statistical deviations of our considered model from  $\Lambda$CDM can be calculated as 
$\Delta\rm{AIC}=\big|\rm{AIC}_{\rm{Model}}-\rm{AIC}_{\Lambda\rm{CDM}}\big|\;,$  and $\Delta\rm{BIC}=\big|\rm{BIC}_{\rm{Model}}-\rm{BIC}_{\Lambda\rm{CDM}}\big|$. If $\Delta\rm{AIC}\leq2$ the model has substantial observational support for the fitted data; if $4\leq\Delta\rm{AIC}\leq7$, the model has less observational support; and if $\Delta\rm{AIC}\geq10$, the model has no observational support. For the case $0\leq\Delta\rm{BIC}\leq2$ it has substantial support; if $2\leq\Delta\rm{BIC}\leq6$, less support; $6\leq\Delta\rm{BIC}\leq10$, mild support; and $\Delta\rm{BIC}>10$, unsupported. For more information on these methods, see e.g.  \cite{liddle2009statistical,szydlowski2015aic,rezaei2021comparison}.  We calculate the above statistical information for the current models and summarize it in Table  \ref{tab:results_95}. The results show that both the CIDE and UCSFDE models yield $\chi^2$ values close to those of  $\Lambda$CDM model across all dataset combinations. Based on the $\Delta\mathrm{AIC}$ values, both models receive substantial statistical support from the data, except when using the \texttt{CMB+  DESI DR2+PP}. For the CIDE model, the value of $\Delta\mathrm{AIC}$ for   \texttt{CMB+  DESI DR2+ PPS}  combinations falls within the range of less support. However, due to its additional parameters (see more \cite{sahlu2025structure}), the CIDE model has mild support by  $\Delta\mathrm{BIC}$, placing it in a regime of weak dataset combinations for  CMB,  \texttt{CMB +   DESI  DR2}, and  CMB +   DESI DR2 + DD. The same scenario has been made for the UCSFDE model for CMB datasets. 

\begin{table*}[htbp]
\centering
\renewcommand{\arraystretch}{1.3}
\caption{Inferred Hubble constant $H_0$ in km/s/Mpc and corresponding tension levels for the CIDE, UCSFDE, and $\Lambda$CDM models in relation to Planck 2018 ($67.4 \pm 0.5$ \citep{aghanim2020planck}) and SH0ES 2022 ($73.04 \pm 1.04$ \citep{riess2021comprehensive}). }
\label{tab:H0_tensions}
\small
\begin{tabular}{l|ccc|ccc|ccc}
\hline
\multirow{2}{*}{Dataset} & \multicolumn{3}{c|}{CIDE} & \multicolumn{3}{c|}{UCSFDE} & \multicolumn{3}{c}{$\Lambda$CDM} \\
 & $H_0$ & Planck & SH0ES & $H_0$ & Planck & SH0ES & $H_0$ & Planck & SH0ES \\
\hline
\texttt{CMB} & $67.76^{+4.17}_{-3.87}$ & $0.09\sigma$ & $1.27\sigma$ & $67.23^{+1.10}_{-1.22}$ & $0.14\sigma$ & $3.73\sigma$ & $67.38^{+0.97}_{-1.00}$ & $0.02\sigma$ & $3.95\sigma$ \\
\texttt{CMB $+$ DESI DR2} & $68.91^{+0.96}_{-0.95}$ & $1.40\sigma$ & $2.93\sigma$ & $68.33^{+0.57}_{-0.61}$ & $1.20\sigma$ & $3.95\sigma$ & $68.41^{+0.62}_{-0.65}$ & $1.25\sigma$ & $3.80\sigma$ \\
CMB $+$ DESI DR2 $+$ DD & $68.65^{+0.76}_{-0.86}$ & $1.32\sigma$ & $3.33\sigma$ & $68.20^{+0.54}_{-0.53}$ & $1.09\sigma$ & $4.15\sigma$ & $68.29^{+0.55}_{-0.52}$ & $1.21\sigma$ & $4.06\sigma$ \\
CMB $+$ DESI DR2 $+$ PP & $68.77^{+0.85}_{-0.83}$ & $1.40\sigma$ & $3.19\sigma$ & $68.23^{+0.54}_{-0.50}$ & $1.15\sigma$ & $4.13\sigma$ & $68.32^{+0.58}_{-0.60}$ & $1.20\sigma$ & $3.95\sigma$ \\
CMB $+$ DESI DR2 $+$ PPS & $69.53^{+0.79}_{-0.83}$ & $2.23\sigma$ & $2.66\sigma$ & $68.67^{+0.57}_{-0.53}$ & $1.71\sigma$ & $3.71\sigma$ & $68.75^{+0.63}_{-0.60}$ & $1.72\sigma$ & $3.55\sigma$ \\
CMB $+$ DESI DR2 $+$ U3 & $68.75^{+0.89}_{-0.96}$ & $1.28\sigma$ & $3.08\sigma$ & $68.23^{+0.64}_{-0.55}$ & $1.07\sigma$ & $4.01\sigma$ & $68.25^{+0.61}_{-0.58}$ & $1.09\sigma$ & $3.99\sigma$ \\
\hline
\end{tabular}
\end{table*}

\begin{figure}[h!]
    \centering
    \includegraphics[width=0.79\linewidth]{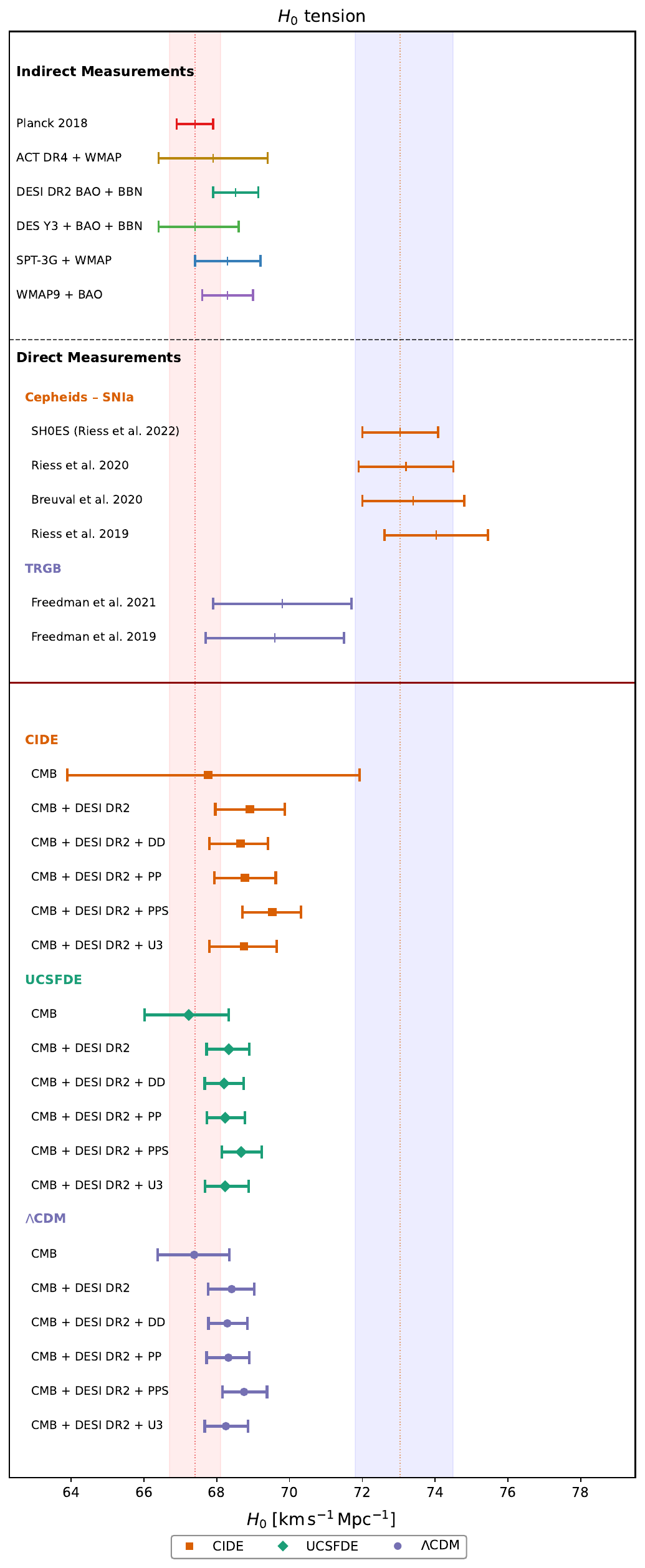}
    \caption{The comparison of the $H_0$  values  (taken from Table~\ref{tab:results_95}) in km/s/Mpc  between  our current models: $\Lambda$CDM, CIDE and UCSFDE with various indirect \cite{aghanim2020planckx,abbott2018dark,addison2018elucidating,aiola2020atacama,ivanov2021cosmological} and direct measurements \cite{riess2022comprehensive,riess2021cosmic,freedman2012carnegie,riess2019large,breuval2020milky}.} 
    \label{fig:H0}
\end{figure}
\begin{figure}[h!]
    \centering
    \includegraphics[width=0.79\linewidth]{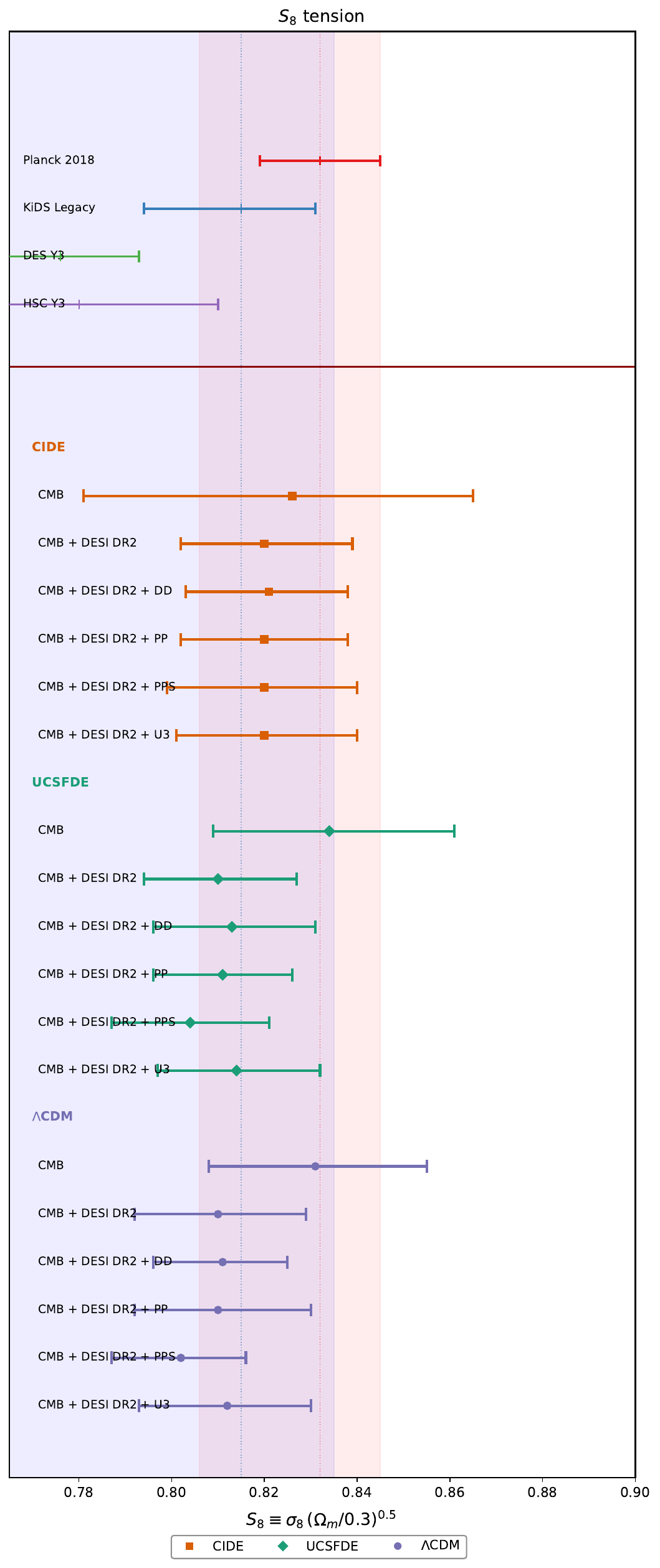}
     \caption{The comparison of the $S_8$  values  (taken from Table~\ref{tab:results_95})  between  our current models: $\Lambda$CDM, CIDE and UCSFDE with various measurements \cite{aghanim2020planck,aghanim2020planck,troxel2018dark,abbott2022dark,asgari2021kids,wright2025kids}.} 
    \label{fig:S8}
\end{figure}

\begin{table*}[htbp]
\centering
\renewcommand{\arraystretch}{1.3}
\caption{The matter clustering  $S_8$  values and corresponding tension levels for the CIDE, UCSFDE, and $\Lambda$CDM models in relation to Planck 2018 ($S_8 = 0.832\pm 0.013$ \citep{aghanim2020planck}),  KiDS Legacy $S_8 = 0.815^{+0.016}_{-0.021}$ \citep{wright2025kids}) and DES Y6 $S_8 = 0.789\pm 0.012$ \cite{sanchez2026dark}. }
\label{tab:S8_tensions}
\small
\begin{tabular}{ll|cccc}
\hline\hline
Model & Dataset & $S_8$ values & Planck 2018 & KiDS Legacy & DES Y6 \\
\hline
\multirow{6}{*}{CIDE} & \texttt{CMB} & $0.826^{+0.039}_{-0.045}$ & $0.14\sigma$ & $0.24\sigma$ & $0.85\sigma$ \\
 & \texttt{CMB $+$ DESI DR2} & $0.820^{+0.019}_{-0.018}$ & $0.53\sigma$ & $0.20\sigma$ & $1.41\sigma$ \\
 & \texttt{ CMB $+$ DESI DR2 $+$ DD }& $0.821^{+0.017}_{-0.018}$ & $0.50\sigma$ & $0.25\sigma$ & $1.51\sigma$ \\
 & \texttt{ CMB $+$ DESI DR2 $+$ PP}& $0.820^{+0.018}_{-0.018}$ & $0.54\sigma$ & $0.21\sigma$ & $1.43\sigma$ \\
 & \texttt{ CMB $+$ DESI DR2 $+$ PPS }& $0.820^{+0.020}_{-0.021}$ & $0.49\sigma$ & $0.19\sigma$ & $1.31\sigma$ \\
 & \texttt{ CMB $+$ DESI DR2 $+$ U3}& $0.820^{+0.020}_{-0.019}$ & $0.51\sigma$ & $0.20\sigma$ & $1.35\sigma$ \\
\hline
\multirow{6}{*}{UCSFDE} & \texttt{CMB} & $0.834^{+0.027}_{-0.025}$ & $0.07\sigma$ & $0.62\sigma$ & $1.57\sigma$ \\
 & CMB $+$ DESI DR2 & $0.810^{+0.017}_{-0.016}$ & $1.05\sigma$ & $0.19\sigma$ & $1.03\sigma$ \\
 & \texttt{ CMB $+$ DESI DR2 $+$ DD }& $0.813^{+0.018}_{-0.017}$ & $0.87\sigma$ & $0.07\sigma$ & $1.13\sigma$ \\
 & \texttt{ CMB $+$ DESI DR2 $+$ PP }& $0.811^{+0.015}_{-0.015}$ & $1.06\sigma$ & $0.15\sigma$ & $1.15\sigma$ \\
 & \texttt{ CMB $+$ DESI DR2 $+$ PPS }& $0.804^{+0.017}_{-0.017}$ & $1.31\sigma$ & $0.41\sigma$ & $0.72\sigma$ \\
 & \texttt{ CMB $+$ DESI DR2 $+$ U3}& $0.814^{+0.018}_{-0.017}$ & $0.83\sigma$ & $0.04\sigma$ & $1.18\sigma$ \\
\hline
\multirow{6}{*}{$\Lambda\text{CDM}$} & \texttt{CMB} & $0.831^{+0.024}_{-0.023}$ & $0.04\sigma$ & $0.56\sigma$ & $1.59\sigma$ \\
 & \texttt{CMB $+$ DESI DR2} & $0.810^{+0.019}_{-0.018}$ & $0.97\sigma$ & $0.18\sigma$ & $0.95\sigma$ \\
 & \texttt{ CMB $+$ DESI DR2 $+$ DD }& $0.811^{+0.014}_{-0.015}$ & $1.08\sigma$ & $0.16\sigma$ & $1.17\sigma$ \\
 & \texttt{ CMB $+$ DESI DR2 $+$ PP }& $0.810^{+0.020}_{-0.018}$ & $0.96\sigma$ & $0.18\sigma$ & $0.93\sigma$ \\
 & \texttt{ CMB $+$ DESI DR2 $+$ PPS }& $0.802^{+0.014}_{-0.015}$ & $1.54\sigma$ & $0.51\sigma$ & $0.69\sigma$ \\
 & \texttt{ CMB $+$ DESI DR2 $+$ U3}& $0.812^{+0.018}_{-0.019}$ & $0.88\sigma$ & $0.11\sigma$ & $1.04\sigma$ \\
\hline\hline
\end{tabular}
\end{table*}

\subsection{$H_0$ and $S_8$ Tensions}
In this section, we aim to investigate whether the CIDE model can potentially alleviate the cosmological tensions ($H_0$ and $S_8$), using the $S_8$ vs. $H_0$ diagram from our MCMC contour plots presented in Fig.~\ref{fig:tension1}.

A well-known discrepancy exists between early-time measurements from Planck 2018 ($H_0 = 67.4 \pm 0.5~\mathrm{km\,s^{-1}\,Mpc^{-1}}$) \citep{aghanim2020planck} and local late-time determinations such as the SH0ES 2022 SNIa measurement ($H_0 = 73.04 \pm 1.04~\mathrm{km\,s^{-1}\,Mpc^{-1}}$) \citep{riess2021comprehensive}, yielding a $5.28\sigma$ tension. Intermediate measurements, such as   DESI DR2 2024 ($H_0 = 68.52 \pm 0.62~\mathrm{km\,s^{-1}\,Mpc^{-1}}$) \citep{adame2025desi}, further highlight the complexity of this discrepancy. This persistent mismatch may stem from unaccounted systematic errors or point toward novel physics beyond the standard $\Lambda$CDM paradigm.

Table~\ref{tab:H0_tensions} summarizes the statistical tension levels for the CIDE, UCSFDE, and $\Lambda$CDM models relative to the Planck 2018 and SH0ES 2022 measurements. Across all dataset combinations, the deviation significance for the CIDE model remains below $2.23\sigma$ relative to Planck 2018 and below $3.33\sigma$ relative to SH0ES 2022. By comparison, the UCSFDE model yields deviations of $\le 1.71\sigma$  with Planck 2018  and $\le 4.15\sigma$  with SH0ES 2022, whereas $\Lambda$CDM exhibits larger discrepancies of $\le 1.72 \sigma$ and $\le  4.06\sigma$, respectively.

Furthermore, Fig. \ref{fig:H0} shows a quantitative comparison of the CIDE and UCSFDE model results with various indirect and direct measurements of $H_0$ in km.s/Mpc. The $H_0$ values from indirect measurements are derived from \texttt{Planck 2018} $ H_0 = 67.4\pm 0.5$ \cite{aghanim2020planck}, \texttt{ DESY1+BAO+BBN}  $H_0 = 67.4^{1.2}_{ 1.0}$ \cite{abbott2018dark}, \texttt{  DESI DR2 DR1 BAO+BBN (2024)},  $H_0 = 68.52\pm 0.62$ \cite{adame2025desi}, \texttt{ACT DR6 + CMB} $H_0 = 67.9\pm 1.5$ \cite{aiola2020atacama}, \texttt{WMAP9+galaxy+$\rm L_{y\alpha}$ BAO} $H_0 = 68.3 \pm 0.7$ \cite{addison2018elucidating}, \texttt{SPT-SZ+ galaxy+BAO} $68.3\pm$ \cite{addison2018elucidating}, \texttt{BOSS+eBOSS FS+BAO+BBN} $H_0 =68.6\pm1.1$ \cite{ivanov2021cosmological}. The direct measurements of the Hubble parameter values use the Cepheid-Type Ia supernova local distance ladder, geometrically calibrated with Cepheids to determine the luminosity of the Type Ia supernova, including the work in  \cite{riess2022comprehensive, riess2021cosmic,freedman2012carnegie,riess2019large,breuval2020milky}. 

Our models' predicted $H_0$ values using combined early- and late-time measurements are shown in Fig.~\ref{fig:H0} and summarized in Table~\ref{tab:H0_tensions}. Notably, these values lie directly in the intermediate regime between local (direct) and early-universe (indirect) measurements. The CIDE model yields tight error bars centered between the two observational extremes. The UCSFDE model provides even tighter error bars, but the model is not an effective bridge, particularly for combinations and does not alleviate the Hubble tension, as discussed above. However, neither model helps alleviate the $H_0$ tension.

Additionally, we draw attention to the tension for the matter clustering parameter, $S_8 \equiv \sigma_8\sqrt{\Omega_m/0.3}$, which quantifies the amplitude of matter fluctuations. A notable tension exists between the early-Universe determination from Planck 2018 ($S_8 = 0.832 \pm 0.013$) \cite{aghanim2020planck} and recent weak lensing/galaxy clustering surveys, such as the DE Survey Year 6 (DES Y6; $S_8 = 0.789 \pm 0.012$) \cite{sanchez2026dark}, which exhibits a $2.4\sigma$–$2.7\sigma$ discrepancy with standard $\Lambda$CDM predictions \cite{PANTOS2026102286}. In contrast, the tension relative to KiDS-Legacy ($S_8 = 0.815^{+0.016}_{-0.021}$) is milder ($< 1\sigma$). Although the systematic origin of the $S_8$ discrepancy remains actively debated, we evaluate our model against these survey constraints to test its potential in mitigating the clustering tension. Table~\ref{tab:S8_tensions} summarizes the statistical deviations ($\sigma$) between our theoretical predictions and these observational datasets. Across all combined datasets, all three models maintain consistency within $< 2\sigma$ for DES Y6 and $< 1\sigma$ for KiDS-Legacy. By comparison,  UCSFDE and standard $\Lambda$CDM  models exhibit larger discrepancies compared with the CIDE model. By such comparison, the CIDE model is a better candidate to mitigate the $S_8$ tension.  

Fig.~\ref{fig:S8} illustrates these matter clustering constraints alongside various observational determinations: Planck 2018 ($S_8 = 0.832 \pm 0.013$) \cite{aghanim2020planck}, DES Y1 ($S_8 = 0.783 \pm 0.025$) \cite{troxel2018dark}, DES Y3 ($S_8 = 0.759^{+0.024}_{-0.021}$) \cite{abbott2022dark}, KiDS-1000 ($S_8 = 0.766 \pm 0.020$) \cite{asgari2021kids}, KiDS-Legacy ($S_8 = 0.815^{+0.016}_{-0.021}$) \cite{wright2025kids}, and DES Y6 ($S_8 = 0.789 \pm 0.012$) \cite{sanchez2026dark}.

Overall, these results demonstrate that the CIDE framework has not effectively alleviated the $H_0$ tension while remaining fully compatible with current matter-clustering measurements and solves the $S_8$ tension.

In the following sections, we present the numerical analysis of the background and the perturbation quantities for further investigation of the effect of the CIDE model on the cosmic history and structure formation together with the standard model of cosmology.  Note that we use the constraining values taken from Table \ref{tab:constraints} from\texttt{ CMB $+$ DESI DR2} datasets for illustration purposes, and this random choice is made to avoid overcrowded lines in the single figure.  

\section{Background quantities}
\begin{figure}[t!]
    \centering
    \includegraphics[width=0.95\linewidth]{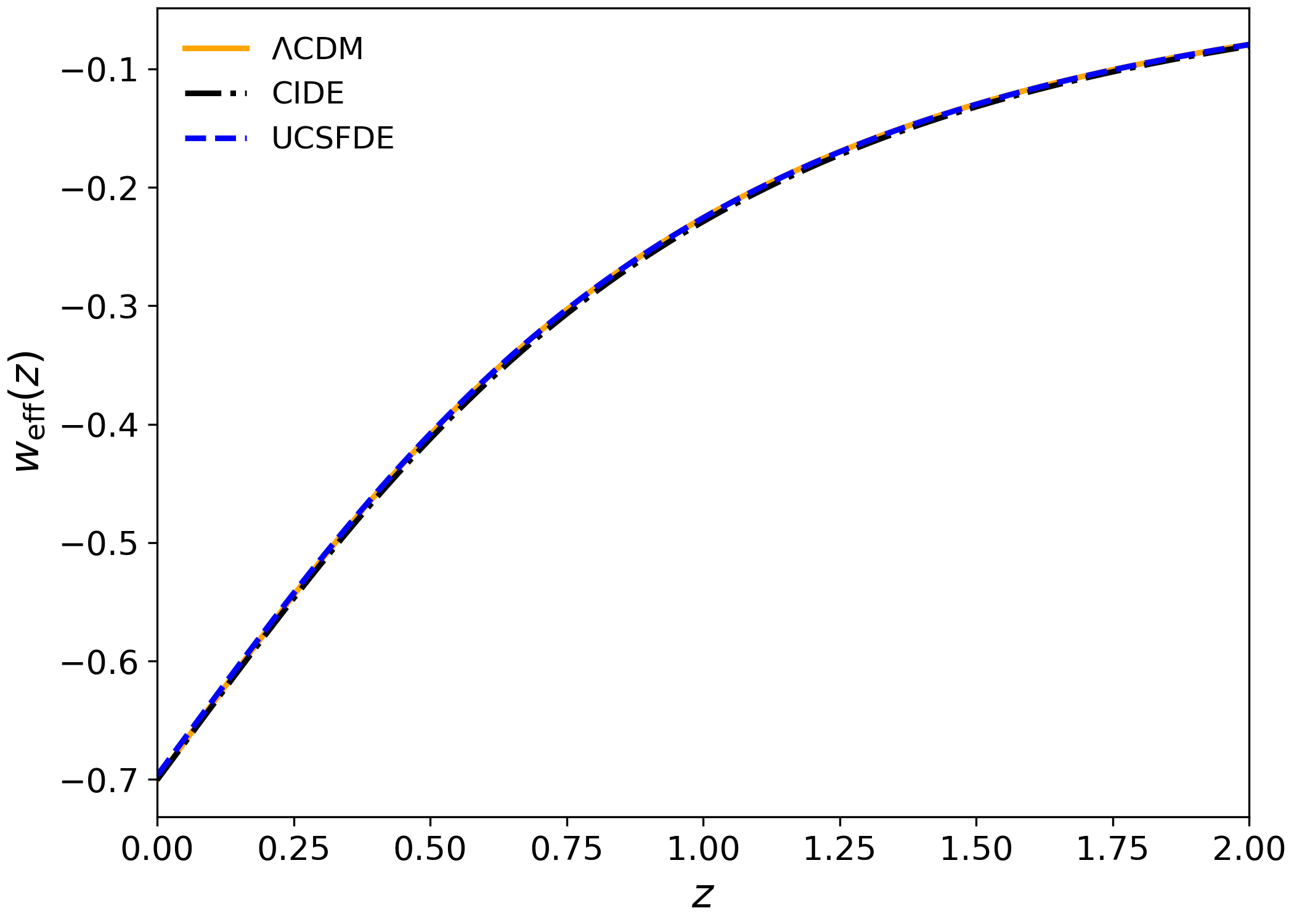}
    \caption{The numerical plots of the effective equation of state parameter for CIDE, UCSFDE, and $\Lambda$CDM models. The values of the constraining parameter are taken from Table \ref{tab:constraints} from the\texttt{ CMB $+$ DESI DR2} datasets. }
    \label{fig:EQ}
\end{figure}

\begin{figure}[t!]
    \centering
    \includegraphics[width=0.95\linewidth]{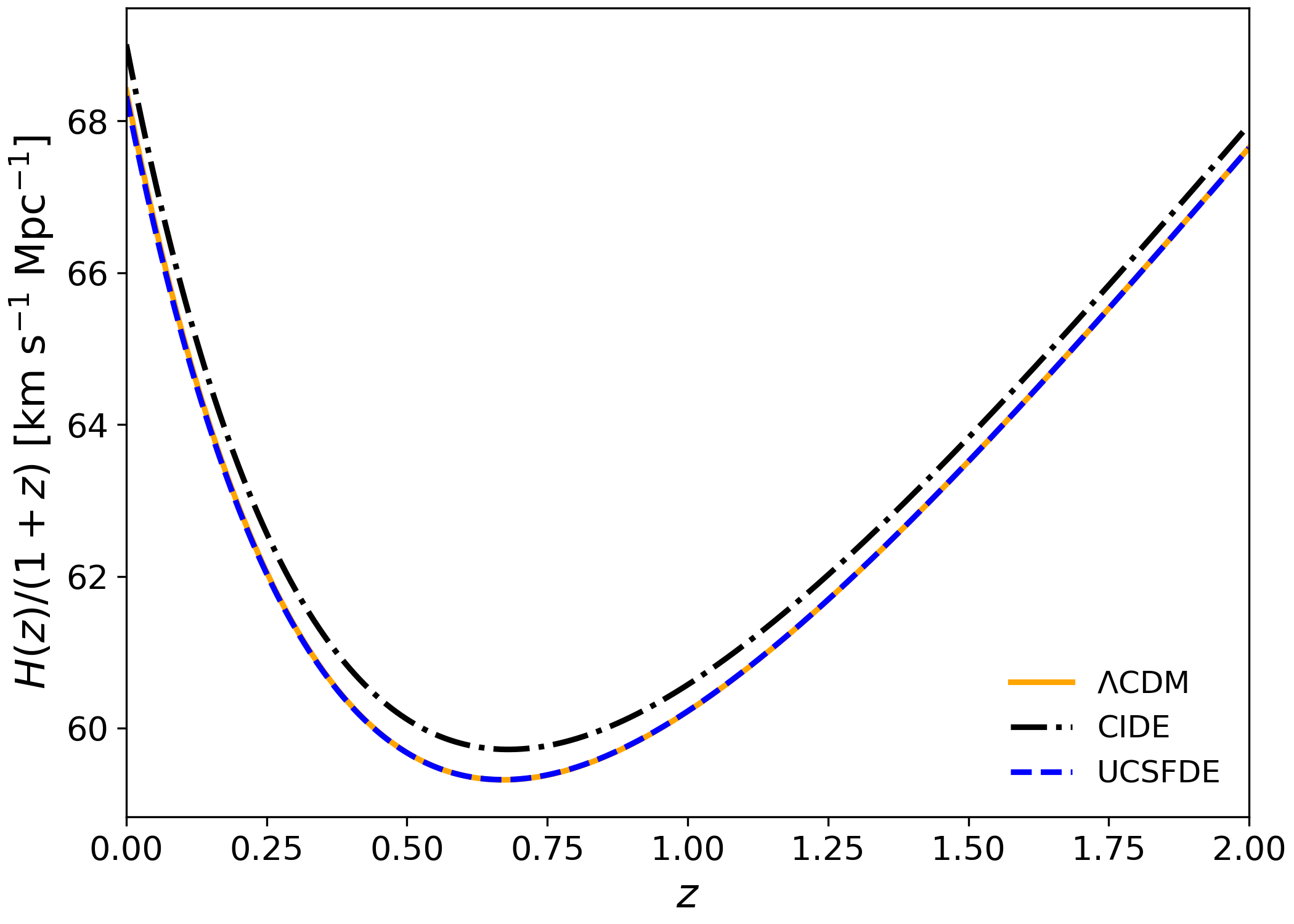}
    \caption{The numerical plots of the rate of the Hubble function $H(z)/(1+z)$ in km/s/Mpc for CIDE, UCSFDE, and $\Lambda$CDM models. The values of the constraining parameter are taken from Table \ref{tab:constraints} from the\texttt{ CMB $+$ DESI DR2} datasets. }
    \label{fig:H0values}
\end{figure}

In this section, we show the background quantities of CIDE, UCSFDE, and the $\Lambda$CDM model together to see how much the scalar field deviates from the standard model of cosmology. For an illustrative case, we choose the effective equation of state parameter and the Hubble functions that show the effect of the scalar field on the cosmic history. 
 We present the numerical results of the effective equation of state parameter \begin{eqnarray}
    w_{eff} = \frac{P_m + P_\phi}{\rho_m + \rho_\phi}\;,
\end{eqnarray}
in Fig \ref{fig:EQ}. From this plot, the CIDE and UCSFDE models explain the cosmic history as the standard model of cosmology. Because of the small coupling parameter $m$, CIDE slightly deviates from $\Lambda$CDM. This deviation is slightly larger in the Hubble function, as shown in Fig. \ref{fig:H0values}, which shows the effect of the CIDE model on the background cosmic expansion history. From both Figs. \ref{fig:EQ} and \ref{fig:H0values}, it is found that the UCSFDE model gives the minimal deviation from the standard model of cosmology. 
\begin{figure}[t!]
     \includegraphics[width=0.95\linewidth]{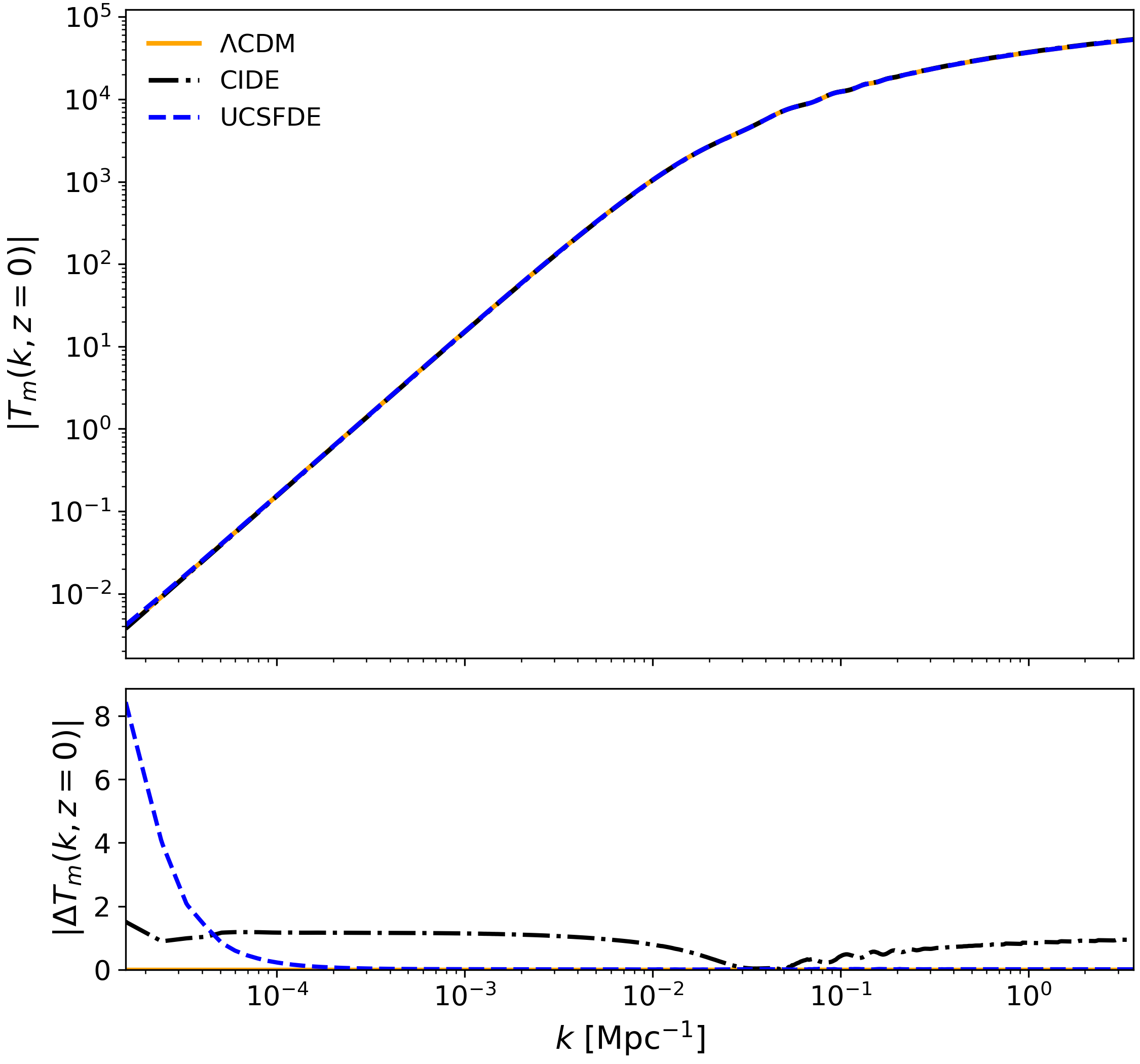}
     \caption{The matter transfer function with the corresponding deviations for CIDE, UCSFDE, and $\Lambda$CDM models. We use the parameter values $n_s$ $\Omega_\mathrm{b} h^2$,$\Omega_\mathrm{c} h^2$,$m$,$H_0$,$S_8 $ for  \texttt{CMB+  DESI DR2} datasets taken from Table \ref{tab:constraints} for all considered models. }
     \label{fig:tk}
 \end{figure}
 \begin{figure*}[t!]
 \includegraphics[width=0.45\linewidth]{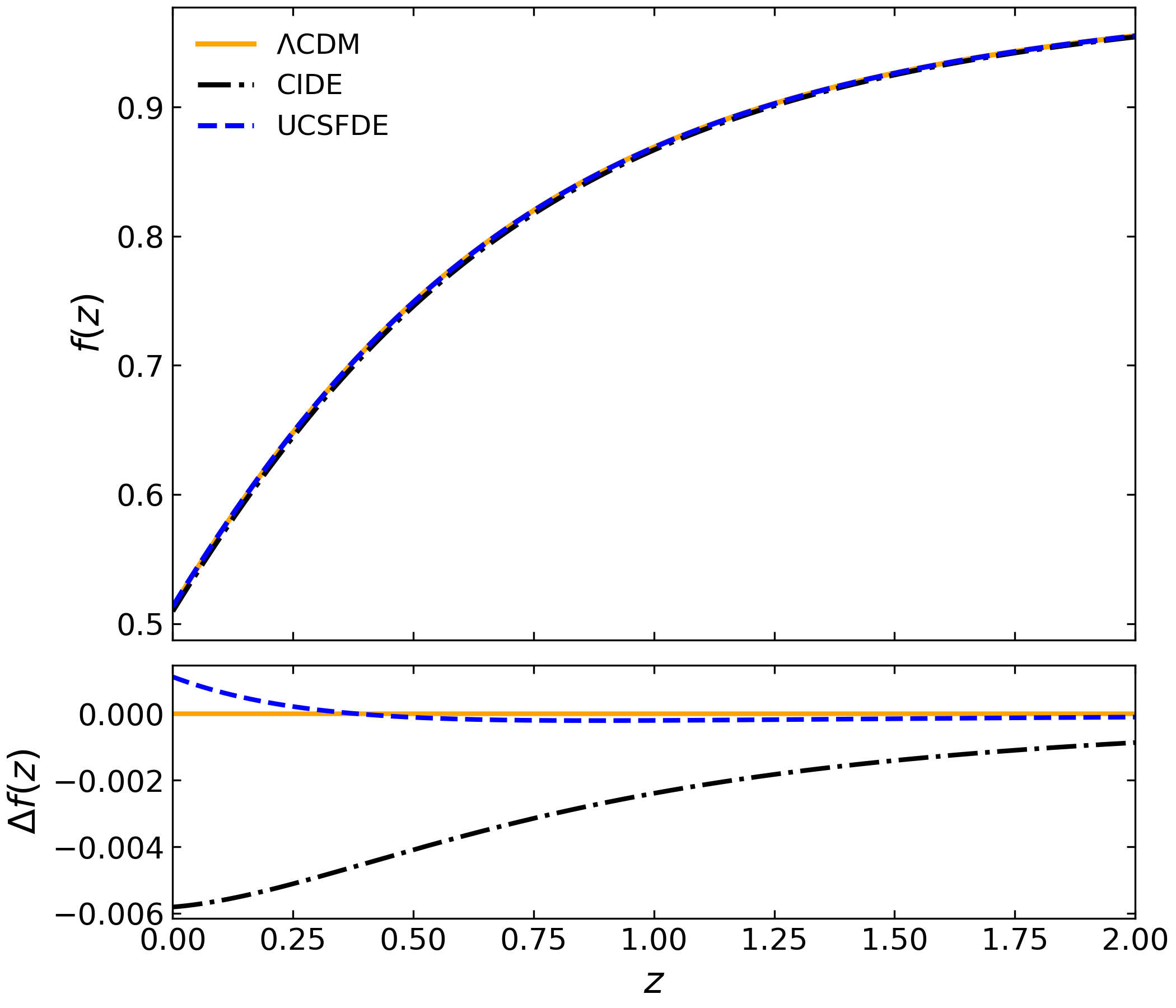}
    \includegraphics[width=0.45\linewidth]{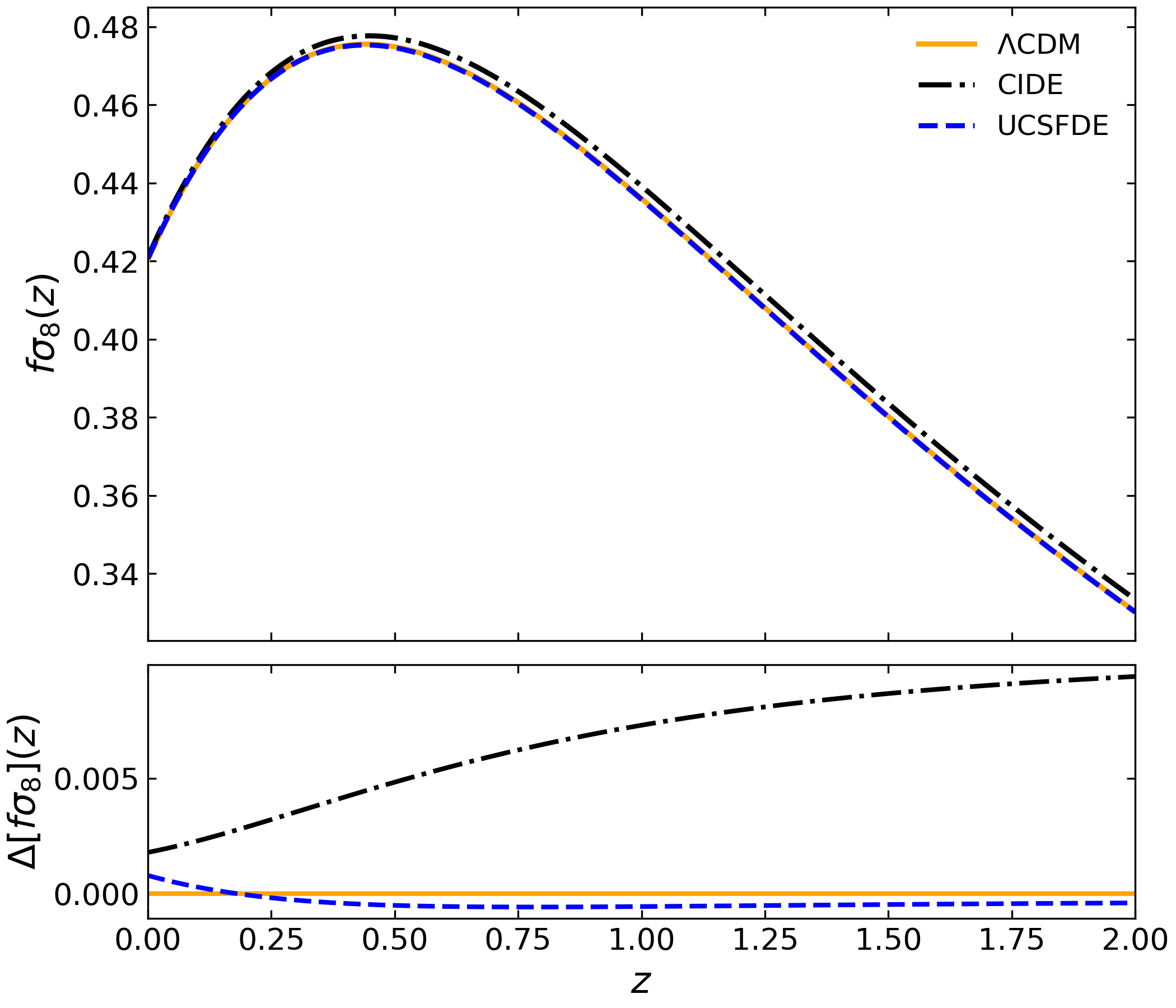}
    \caption{\texttt{Left panel:} The diagram of the growth rate  $f(z)$ of CIDE and UCSFDE compared with the $\Lambda$CDM models. \texttt{Right panel:} The diagram of the RSD $f\sigma_8(z)$ of CIDE and UCSFDE compared with the $\Lambda$CDM models. We use the parameter values $n_s$ $\Omega_\mathrm{b} h^2$,$\Omega_\mathrm{c} h^2$,$m$,$H_0$,$S_8 $ for  \texttt{CMB+  DESI DR2} datasets taken from Table \ref{tab:constraints} for all considered models. }
    \label{fig:placeholderf}
\end{figure*}
 \begin{figure}[t!]
    \includegraphics[width=1.0\linewidth]{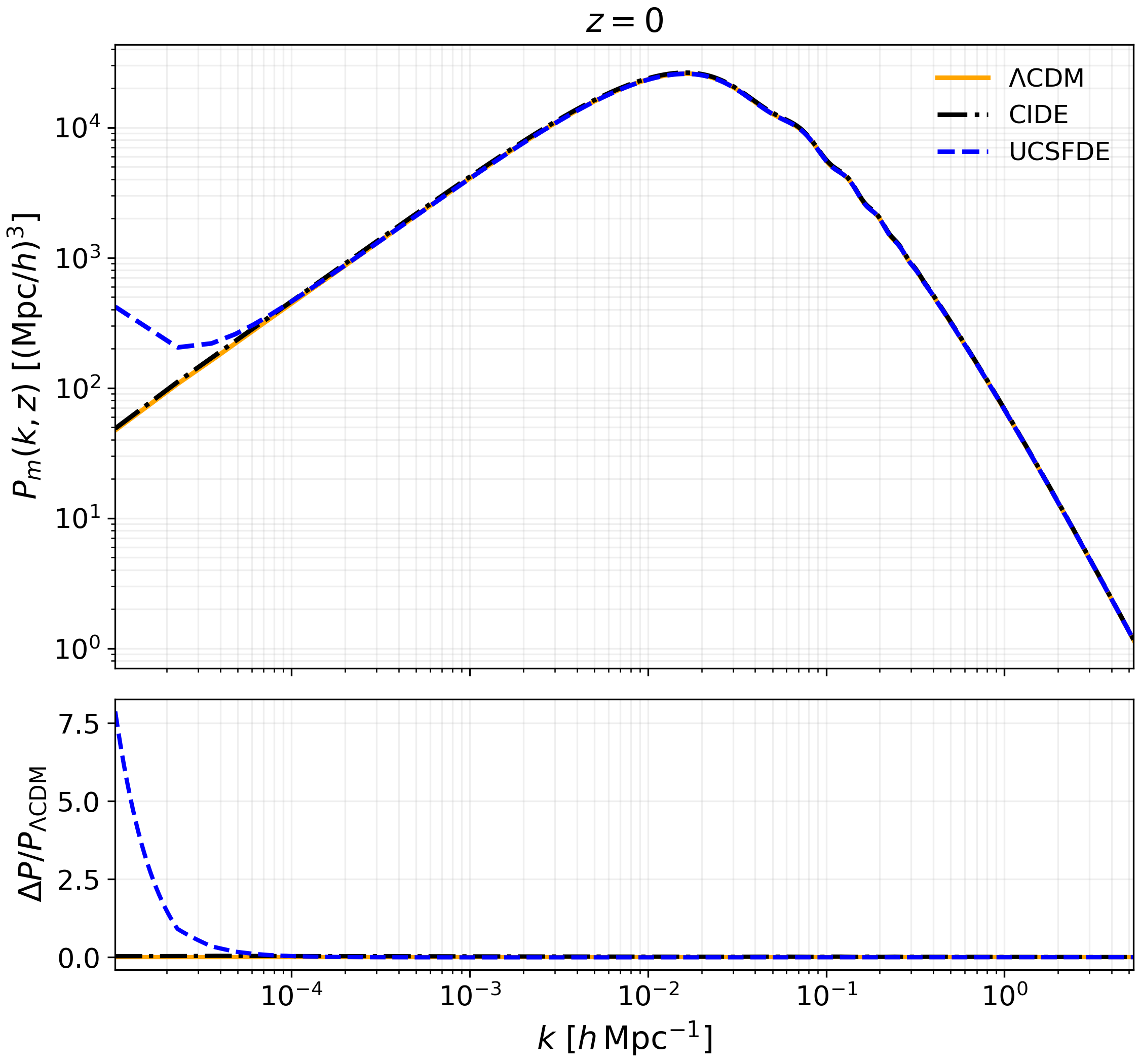}
    \caption{\texttt{Upper panel}: the matter  power spectrum $P_m(k, \;z = 0) $ for   CIDE and UCSFDE against with $\Lambda$CDM. \texttt{Bottom panel}: the deviation of matter power spectrum $\Delta P_m(k, \;z = 0) $ between CIDE, UCSFDE  and $\Lambda$CDM models.   We use the parameter values $n_s$ $\Omega_\mathrm{b} h^2$,$\Omega_\mathrm{c} h^2$,$m$,$H_0$,$S_8 $ for  \texttt{CMB+  DESI DR2} datasets taken from Table \ref{tab:constraints} for all considered models.}
    \label{fig:placeholdermatter}
\end{figure}
\begin{figure}
    \centering
    \includegraphics[width=0.85\linewidth]{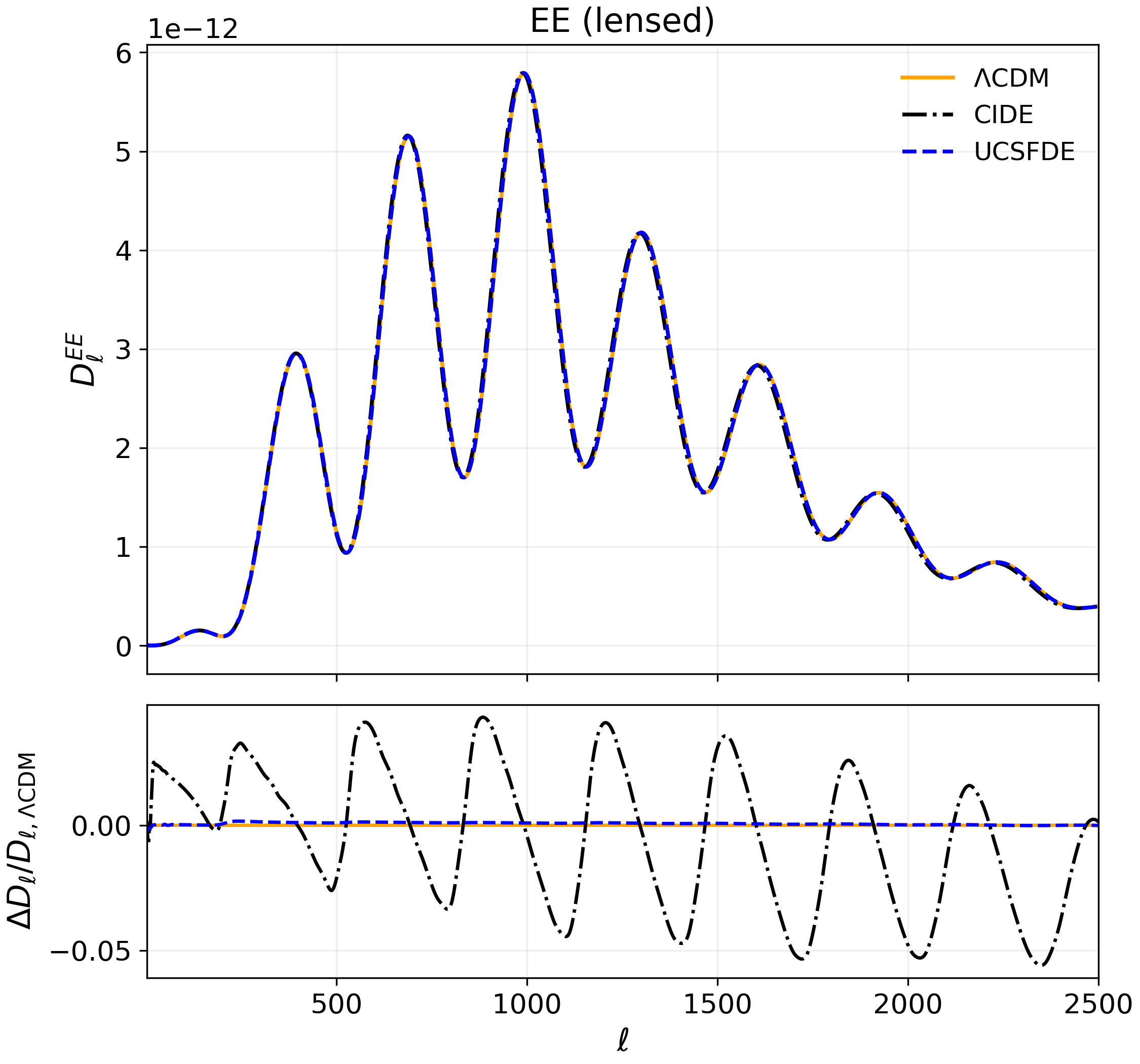}
     \includegraphics[width=0.85\linewidth]{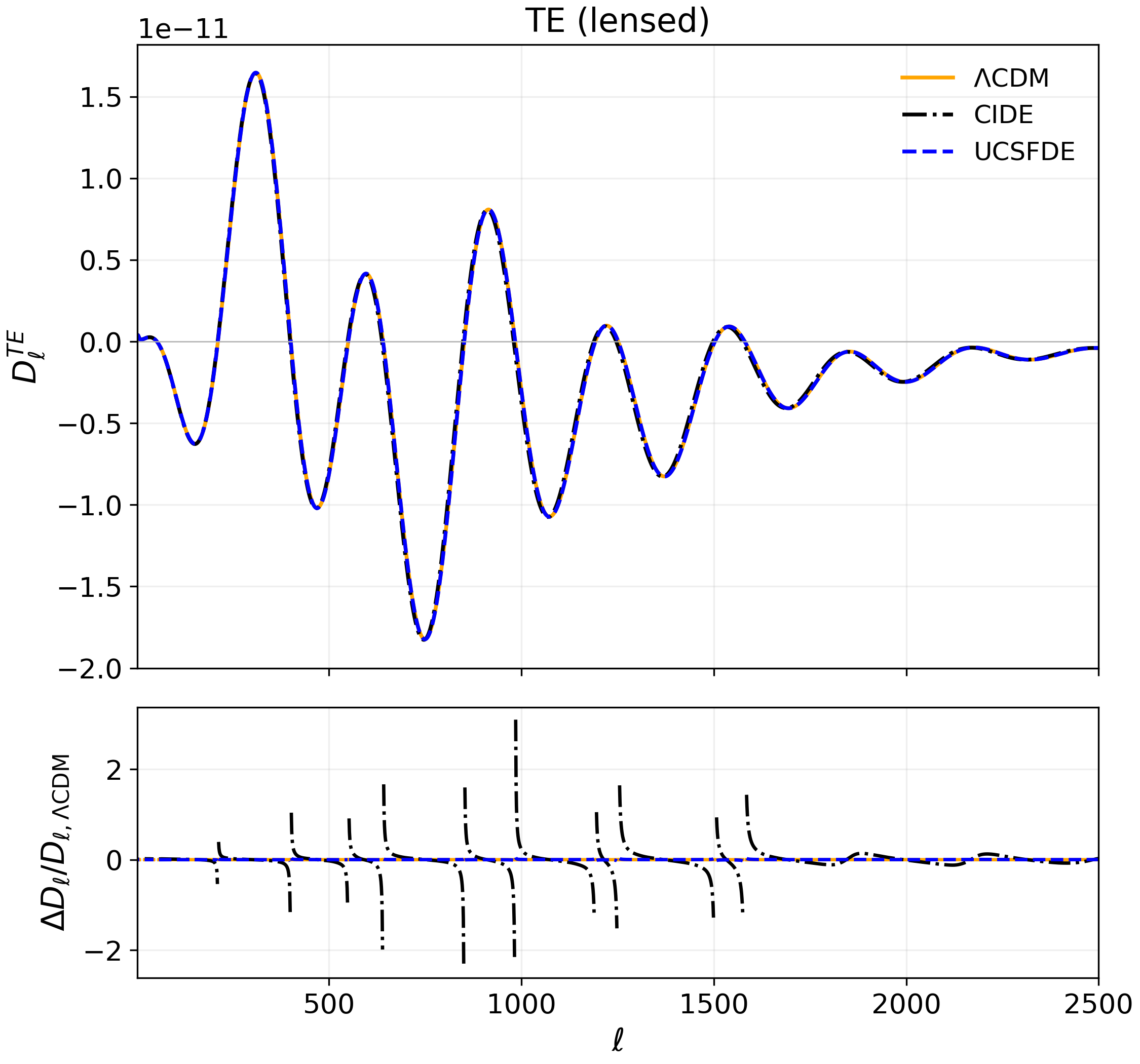}
      \includegraphics[width=0.85\linewidth]{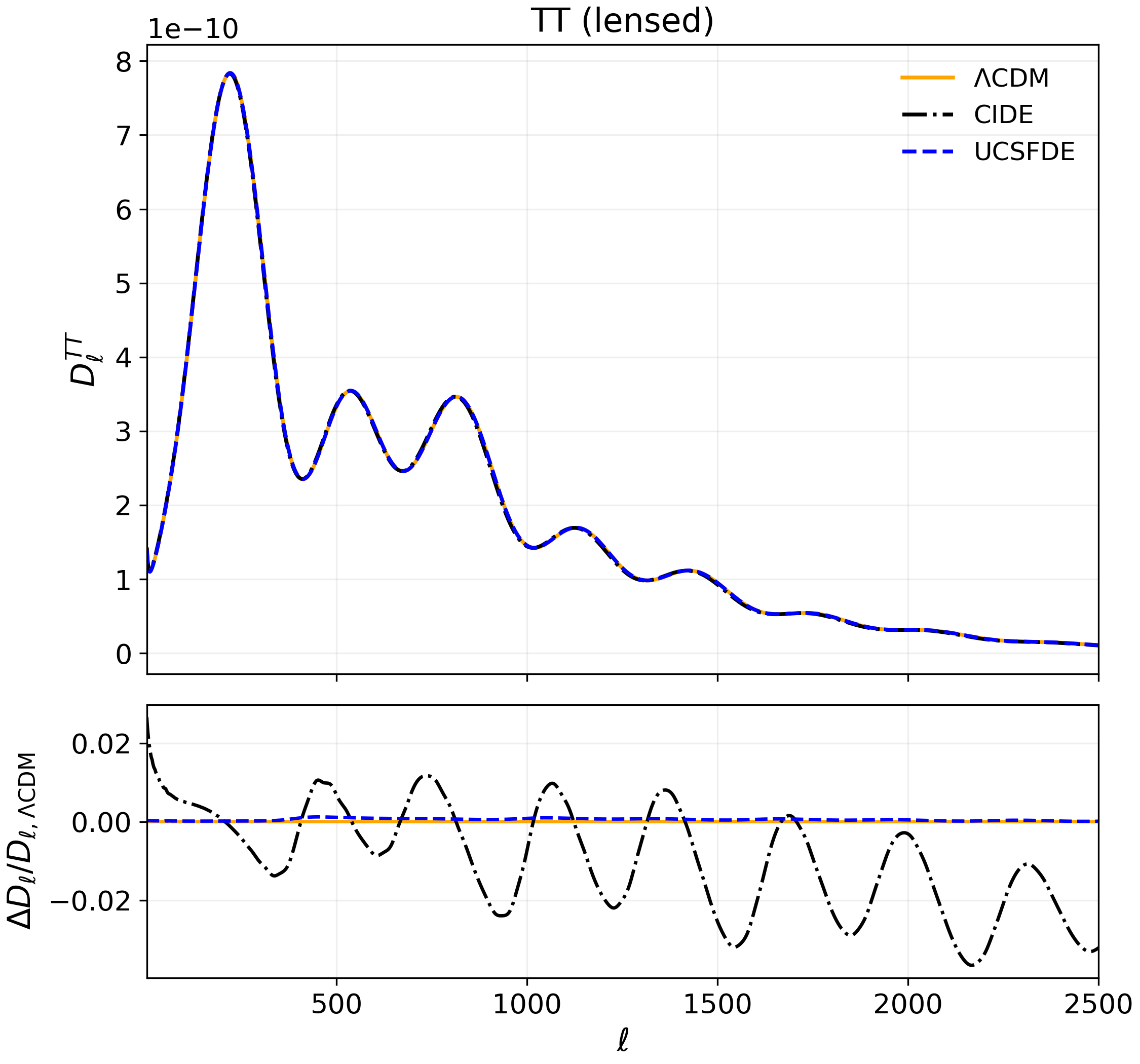}
    \caption{\texttt{Upper, middle, and bottom panels} represent the CMB angular power spectra for the CIDE and UCSFDE models against the $\Lambda$CDM model for TT, EE, and TE polarizations across the multipoles $2\leq l\leq 2500$ respectively, with the corresponding fractional deviation for each polarization.  We use the parameter values $n_s$ $\Omega_\mathrm{b} h^2$,$\Omega_\mathrm{c} h^2$,$m$,$H_0$,$S_8 $ for  \texttt{CMB+  DESI DR2} datasets taken from Table \ref{tab:constraints} for all considered models. }
    \label{fig:placeholderCMB}
\end{figure}

\subsection{Structure growth}
We now focus on the effect of CIDE models on the growth of cosmic structure. The total clustering matter density (baryonic and DM) contrast is defined as \cite{villa2020international,pourtsidou2013models,gomez2018density}:
\begin{eqnarray}
    \delta_m(k,z) = \frac{\rho_b(z)\delta_b(k,z) + \rho_{DM}(z)\delta_{DM}(k,z)}{\rho_b(z)+\rho_{DM}(z)}\;.
\end{eqnarray}
This equation describes the total gravitational effect on the formation of structure in the universe by neglecting DE in the non-clustering components.  In the same manner, the total transfer function, $  T_m(k,z)$, can be expressed as
\begin{eqnarray}
    T_m(k,z) = \Bigg|\frac{\rho_b(z)T_{\delta_b}(k,z) + \rho_{DM}(z)T_{\delta_{DM}}(k,z)}{\rho_b(z)+\rho_{DM}(z)}\Bigg |\;,
\end{eqnarray}
where $T_{\delta_b}(k,z)$ and $T_{\delta_{DM}}(k,z)$   represent the transfer functions for baryonic and DM fluids, respectively. The numerical results of the transfer function for the present day $z = 0$ are shown in Fig. \ref{fig:tk} with CIDE and UCSFDE in comparison of $\Lambda$CDM models. The  fractional deviations, $\Delta T_m(k) $, are represented as
\begin{eqnarray}
   \Delta T_m(k) = \frac{T^{ \rm {CIDE}, \rm UCSFDE}_m(k) - T^{ \rm \Lambda CDM}_m(k) }{T^{\rm \Lambda CDM}_m(k)} \;.
\end{eqnarray}
and plotted on the bottom panel of Fig. \ref{fig:tk}. 
The deviation  $\Delta T_m(k)$ quantifies the linear matter perturbations, and this plot shows values greater than zero across the entire range of wavenumber $k$. This indicates that both the CIDE and UCSFDE models enhance the structure growth.  Particularly at large scales particularly around $k = 10^{-4}$, the UCSFDE model has significant enhancement,  this model has behaves true for the matter power spectrum fluctuations discuss below. 

For a dimensionless primordial power spectrum $\mathcal{P}*\mathcal{R}(k)$, the matter power spectrum can be expressed as
\begin{equation}
    P_m(k,z) = \frac{2\pi^2}{k^3}\mathcal{P}*\mathcal{R}(k) T^2_m(k,z)\;.
\end{equation}
The amplitude of the matter density contrast that quantifies the spatial variance can also be expressed in terms of the smoothed physical scale radius $R$, the matter power spectrum $P_m(k,z)$ across all spatial scales $k$, and the window function $W(kR)$ \cite{press1974formation,bardeen1986statistics,eisenstein1998baryonic,smith2003stable}. The variance of the matter density field smoothed over a radius $R$ is
\begin{eqnarray}
    \sigma^2(R,z) = \frac{1}{2\pi^2}\int_0^\infty k^2 P_m(k,z) W^2(kR)dk
\end{eqnarray}
where $W(kR)$ is the spherical top-hat window function
\begin{equation}
   W(kR) = 3\frac{\sin(kR)-kR\cos(kR)}{(kR)^3}\;.
\end{equation}
The linear matter variance in a spherical comoving radius $R_8 = 8\;h^{-1}\mathrm{Mpc}$ becomes
\begin{equation}\label{sigma8}
    \sigma^2_8(z) =  \frac{1}{2\pi^2}\int_0^\infty k^2 P_m(k,z) W^2(kR_8)dk
\end{equation}
From Eq. \ref{sigma8}, $\sigma_8(z=0)\equiv\sigma_8$ is one of the primary cosmological parameters that can be constrained through \texttt{MCMC} simulations and used to normalize the matter power spectrum. The growth factor is assumed to be scale-independent over the range of scales considered and relates the matter power spectrum as
\begin{equation}
    P_m(k,z) = D^2(z) P_m(k,0)\;,
\end{equation}
where the normalized growth factor is $$D(z)=\delta_m(z)/\delta_m(z=0)\;.$$ The growth rate function is
\begin{equation}
    f(z) = \frac{d\ln D}{d\ln a}.
\end{equation}

The combination of the growth rate with the amplitude of the matter power spectrum $\sigma_8(z)$  is $f\sigma_8(z) $, one of the key cosmological probes measured by galaxy redshift surveys through redshift-space distortions (RSD), where the peculiar velocities of galaxies are distorted along the line of sight. RSD measurements contain the degeneracy inherent in the measurement of $f(z)$ and $\sigma_8(z)$, and can be expressed as
\begin{eqnarray}
    f\sigma_8(z) = \sigma_{8}(z)f(z)\;.
\end{eqnarray}
The numerical results of $f(z)$ with the corresponding deviation $\Delta f = \frac{f^{\rm model} -f^{\rm \Lambda CDM}}{f^{\rm \Lambda CDM}}$,  are shown in the \texttt{Left panel} of Fig. \ref{fig:placeholderf}. More specifically, it shows how fast the matter overdensity collapses under gravity against the background expansion by tracking the inhomogeneous linear perturbations.  The results also clarify that the impact of interacting DE offers a different strategy for replicating the growth history of $\Lambda$CDM while providing crucial information about the formation of structures. The product of the growth rate and the amplitude of matter fluctuations is also displayed by the  $f\sigma_8(z)$ diagram and its corresponding fractional deviation  $\Delta f\sigma_8 = \frac{f\sigma_8^{\rm model} -f\sigma_8^{\rm \Lambda CDM}}{f\sigma_8^{\rm \Lambda CDM}}$,  in the  \texttt{Right panel} of Fig. \ref{fig:placeholderf}.  These figures illustrate how the conformal interacting dark coupling affects late-time matter clustering and verify that the growth predictions of our CIDE have enhanced the growth of structure compared to UCSFDE and $\Lambda$CDM models. Particularly, our current models are more sensitive to RSD than to the growth rate. 

We also present the matter power spectrum $P_m({k,z =0})$ in Fig. \ref{fig:placeholdermatter} in the CIDE models that encodes the statistical distributions of the cosmological matter density fluctuations as a function of the wave number $k$. The clear deviation of the matter power spectrum  between the conformal CIDE and $\Lambda$CDM models, defined as
$$ \Delta P_m(k, \;z) = \frac{P^{\text{model}}_m(k,z) - P^{\Lambda\text{CDM}}_m(k, \;z)}{P^{\Lambda\text{CDM}}_m(k, \;z)}\;. $$  is shown in the bottom panel of  Fig. \ref{fig:placeholdermatter}. This deviation makes it evident that oscillation curves of the matter spectrum, $\Delta P_m(k,z)$, due to BAO-like wiggles, were more clustered at scales $k \lesssim  10^{-4} h\,\mathrm{Mpc}^{-1}$, where the current model enhances the matter fluctuations compared to $\Lambda$CDM. In contrast,  at scales $k \gtrsim  10^{-4} h\,\mathrm{Mpc}^{-1}$, we clearly see that the CIDE and UCSFDE models produce a monotonically suppressed matter fluctuation compared to $\Lambda$CDM due to the dominant flow of energy from the DM to the DE component. 

We also extended the work to compute the impact of the interacting conformal dark sectors on the CMB anisotropy power spectrum as presented in Fig. \ref{fig:placeholderCMB}. The left-hand side of this plot shows the angular power spectrum 
$$ D^{\rm TT, EE, TE}_l = {l}({l}+1) \frac{C^{\rm TT, EE, TE}_{l}}{2\pi}\;,$$ for the temperature TT, E-mode polarization (EE), and temperature-polarization cross-correlation (TE) across the multipoles $2\leq l\leq 2500$; and the right-hand side of this plot also shows the fractional deviation for each polarization:
$$ \Delta  D_l^{TT,EE,TE} =  \frac{C^{\rm TT, EE, TE}_l(\rm model) -C^{\rm TT, EE, TE}_l(\rm \Lambda CDM)}{C^{\rm TT, EE, TE}_l(\rm \Lambda CDM)}\;, $$
relative to the baseline $\Lambda$CDM model. From these fractional deviations, we clearly see that the TT polarization of the considered models significantly deviates from $\Lambda$CDM at low multipoles, where the  CIDE dynamics alter the late-time integrated Sachs-Wolfe (ISW) effect through the gravitational potential decay. On the other hand, at higher multipoles, the EE and TE polarization spectra exhibit sub-percent peak position shifts and amplitude variations, primarily driven by the effect of the  CIDE on the angular diameter distance $d_A(z)$ to last scattering under coupled scalar field.

\section{Conclusion}\label{section 4}

In this work we have investigated the interaction between the dark sectors, $Q = -\frac{C_\phi(\phi)}{2C(\phi)}\rho_{\rm DM}$, in a cosmological model introducing a scalar field coupled to DM via a conformal transformation. This interaction is derived from a Lagrangian, rather than by assuming a phenomenological interaction term.  We used the modified \texttt{CLASS} in the Newtonian gauge to compute the background cosmological evolution as well as the perturbative thermodynamic evolution, as shown in equations \eqref{theta} and \eqref{XX}. The parameters were constrained using \texttt{COBAYA}. We systematically incorporated the scalar field potential approaches in the \texttt{CLASS} code, using the exponential Albrecht-Skordis potential with the field defined in Planck mass units $M_{pl} =1$.  The preliminary conditions are $\phi_{\rm in} = 100$ and $\phi'_{\rm in} = 0$, without implementing the attractor initial conditions. The scalar field functions as a dark energy density, whereas the contribution of the dark energy fluid and the cosmological constant is set to zero. \texttt{CLASS} uses its shooting technique to find $\Lambda$ at each measured cosmological parameter point in order to meet $\Omega_{\phi} = 1- \sum_i \Omega_i$, where $i$ is the fluid species under consideration.

The analysis of model constraints has been performed using a combination of early- and late-time measurements, presented in Table \ref{tab:constraints}, by considering the dynamical nature of DE, which is highlighted by the two  DE scenario models, CIDE and UCSFDE (see Figs.  \ref{fig:placeholder1} and \ref{fig:placeholder2} for the posterior distributions of the cosmological parameters for the CIDE  and UCSFDE models). The best-fit value of the coupling parameter is $m > 0$ at the 95\% confidence level; the observational constraints consistently support a positive conformal coupling, suggesting energy transfer from DM to DE. From the same Table \ref{tab:results_95}, we found that across all data combinations. Whether cosmic acceleration arises from a cosmological constant or a dynamical component is a central goal of cosmology and supports the recent findings on the nature of DE by Ref. \cite{wang2025did,gu2025dynamical}. Table \ref{tab:results_95} summarizes the statistical validations of these models, and the results show that the CIDE model gives lower $\chi^2$ values across all dataset combinations, indicating a better statistical fit than $\Lambda$CDM. The data provide significant statistical support for both models based on $\Delta\mathrm{AIC}$ values, except for the \texttt{CMB+  DESI DR2+PP} and \texttt{CMB+  DESI DR2+PPS} combinations. The CIDE model exhibits moderate support with observation by $\Delta\mathrm{BIC}$ for its new parameter $m$, resulting in not completely rejected statistical support across all dataset combinations.

We also addressed the cosmological tensions associated with $H_0$ and $S_8$. As presented in Table~\ref{tab:H0_tensions}, we evaluated the statistical tension ($\sigma$) between our constrained parameters and both indirect measurements from Planck 2018 ($H_0 = 67.4 \pm 0.5~\mathrm{km\,s^{-1}\,Mpc^{-1}}$) and local direct determinations from SH0ES 2022 ($H_0 = 73.04 \pm 1.04~\mathrm{km\,s^{-1}\,Mpc^{-1}}$). Across all dataset combinations, the CIDE model maintains deviations below $2.23\sigma$ relative to Planck 2018 and $\le 3.33\sigma$ relative to SH0ES 2022; the UCSFDE model further reduces these discrepancies to $\le 1.71\sigma$ (Planck 2018) and $\le 4.15\sigma$ (SH0ES 2022), whereas $\Lambda$CDM exhibits larger tensions reaching up to $1.71\sigma$ and $4.06\sigma$, respectively. Fig.~\ref{fig:H0} compares CIDE and UCSFDE model fits with indirect and direct measurements of $H_0$ in km/s/Mpc, derived from indirect observations. Notably, these numbers fall squarely in the middle of the local (direct) and early-universe (indirect) observations. While the CIDE model gives narrow error bars centered between the two observational extremes, the UCSFDE model provides an even more effective bridge; thereby, a conformal interacting dark sector is not a fully possible way to relax the $H_0$ tension compared with the existing tension between the direct and indirect measurements. 

Similarly, we reported the $S_8$ tensions in Table \ref{tab:S8_tensions}, comparing the models' values with Planck 2018, KiDS-Legacy, and DESY Y6 measurements, since the systematic debate is ongoing about resolving the $S_8$ tension between KiDS-Legacy and DESY Y6 measurements. From this analysis, we found that both the CIDE and UCSFDE models maintain consistency within $< 2\sigma$ for DES Y6 and $< 1\sigma$ for KiDS-Legacy; for further details, see Fig.~\ref{fig:S8}. 

In addition to the statistical validations of the model and the discussion of its implications for the cosmological tensions, we explored the model's impact on structure growth through the numerical analysis of perturbative thermodynamic quantities. To do so, we presented the numerical results of the total matter transfer functions $T_m(k, z = 0)$ in Fig. \ref{fig:tk} with the corresponding fractional deviation between the CIDE and UCSFDE relative to $\Lambda$CDM. The results showed  $\Delta T (k)> 0$ across all ranges of the wavenumber $\kappa$. Particularly, the UCSFDE model enhanced the matter clustering compared with the $ \Lambda$CDM model. The same scenarios have been noticed for the fractional variations of the matter power spectrum, $\Delta P_m(k,z)$, which display characteristic BAO wiggles and reveal an enhancement of matter fluctuations at small values of  wavenumber range  ($k \gtrsim 10^{-4}~h\,\mathrm{Mpc^{-1}}$) relative to $\Lambda$CDM, see Fig. \ref{fig:placeholdermatter}.  Fig. \ref{fig:placeholderf}  presents the growth rate  $f(z)$ (Left panel) and the RSD  $f\sigma(z)$ (Right panel).In Fig. \ref{fig:placeholderCMB} the CMB angular power spectra for TT, EE, and TE polarizations across the multipoles $2\leq l\leq 2500$ with the corresponding fractional deviations for CIDE and UCSFDE are compared with the $\Lambda$CDM models.  At low multipoles, the interacting DE dynamics modify the late-time integrated Sachs-Wolfe (ISW) effect via the decay of gravitational potentials. This is reflected in the fractional deviations of the CMB temperature (TT) power spectrum, which displays significant departures from $\Lambda$CDM, particularly for the UCSFDE model. At higher multipoles, the EE polarization and TE cross-correlation spectra exhibit only sub-percent peak shifts and amplitude variations, primarily driven by the modified angular diameter distance $d_A(z)$ to the last scattering surface.

Overall, our results confirm that conformally coupled dark energy models offer a robust statistical alternative to $\Lambda$CDM, providing improved $\chi^2$ fits and preserving consistency with $S_8$ clustering constraints. Remarkably, this variant efficiently bridges early- and late-time Universe probes, reducing the $S_8$ tension to below $1.51\sigma$ without sacrificing agreement with CMB observations, while only mildly alleviating the $H_0$ tension.  This work highlights the power of non-minimal dark-sector couplings in challenging the standard paradigm's assumption of separate dark-sector evolution, reinforcing the need to keep studying interacting dark energy physics in future observational analyses. 
\section*{Acknowledgments}
SS thanks the Department of Theoretical Physics at the University of Valencia (Spain) and the Department of Theoretical Physics at the Complutense University of Madrid (Spain) for their kind hospitality during the early phases of the elaboration of this manuscript.  Rethabile Thubisi was initially involved in earlier aspects of the work, and her contribution is duly acknowledged. 
 This work is supported by the Spanish National Grants PID2022-138607NB-I00, PID2023-149560NB-C21,  CNS2024-154444, and the Severo Ochoa Excellence Grant CEX2023-001292-S, funded by MICIU/AEI/10.13039/501100011033 (“ERDF A way of making Europe”, “PGC Generacion de Conocimiento”) and FEDER, UE.  The authors also acknowledge financial support from the project i-COOPB23096 (funded by CSIC). 

\bibliography{biblio}  


\end{document}